\documentclass[aps,prc,noshowpacs,nofootinbib,twocolumn]{revtex4-1}

\usepackage{epsfig}
\usepackage{graphicx,amsmath,latexsym}
\usepackage{color}
\newcommand\ba{\begin{eqnarray}}
\newcommand\ea{\end{eqnarray}}
\newcommand{\be}{\begin{equation}}
\newcommand{\ee}{\end{equation}}
\newcommand{\bas}{\begin{eqnarray*}}
\newcommand{\eas}{\end{eqnarray*}}
\begin{document}
\title{\bf \large Fractals and log-periodic corrections applied to masses\\
and energy levels of several nuclei}

\author{B. Tatischeff$^{1,2}$\\
$^{1}$Univ Paris-Sud, IPNO, UMR-8608, Orsay, F-91405\\
$^{2}$CNRS/IN2P3, Orsay, F-91405}
\thanks{tati@ipno.in2p3.fr}


\vspace*{1cm}
\begin{abstract}
A contribution is presented to the application of fractal properties and log-periodic corrections to the masses of several nuclei (isotopes or isotones), and to the energy levels of some nuclei. The fractal parameters $\alpha$ and $\lambda$ are not randomly distributed, but take a small number of values, common also with the values extracted previously from fractal distributions of quark, lepton, and hadronic masses. Several masses of still unobserved nuclei are tentatively predicted.
\end{abstract}
\maketitle
\section{Introduction}
The powerful concept of fractals \cite{mandelbrot}, has been applied to the study of a large number of different fields (see for example \cite{nottale}). 
However, up to now, to my knowledge, only one attempt was done
to look whether the fractal properties were appropriate to describe, at least partly, the many properties of nuclear physics. The hadronic, lepton, and gauge boson masses,  have been already discussed within a fractal scaling model of a chain system of quantum harmonic oscillators
\cite{mueller}. It was shown there, that "Scaling is a fundamental property of any natural oscillation process". The author interpreted the elementary particle masses as being proton resonances. He uses a different model but also on the logarithmic scale. This approach is different from our approach. 

Several particle physics ratios, and strength ratios, were discussed within a "Higgs free symmetry breaking from critical behavior near dimension four" model \cite{goldfain}.

Before starting the present study, several steps were elaborated:\\
 -  first, several relations between nearly all elementary particle masses were found
\cite{arxiv1005}. The masses of quarks, gauge bosons, and leptons  were reproduced, starting for each species with the lowest mass, using nucleon and pion masses, and fine structure constant value;\\
 - then, it was shown that the fractal properties agree to describe the fundamental force coupling constants, the atomic energies, and the elementary particle masses
 \cite{arxiv1104}. This agreement was obtained using the discrete-scale invariance model (DSI) \cite{sornette} \cite{nottale1};\\
 - finally the fractal properties were applied with success to the hadronic masses
 \cite{frachadron}. It was found that the parameters used to fit the mass ratios between all adjacent masses of a given species, display the same distribution for all hadronic species.\\ 

The very large number of properties concerning nuclear physics, asks for an attempt to rely these properties to each others, as much as possible,
in order to reduce the number of free values. These properties are the nuclear masses, their energy level masses, the total and partial widths for all excited levels, the quantum numbers .... Below, we will discuss some of these properties within the fractal description, precisely to look for eventual common descriptions.\\   

\subsection{The fractal characteristic of hadronic masses}
The fractal concept stipulates that the same physical laws apply for different scales of a given physics \cite{note}. We summarize here briefly the concept of continuous and discrete scale invariances transcribing the developpements of D. Sornette \cite{sornette} and L. Nottale \cite{nottale}. 
The concept of {\it continuous} scale invariance is defined in the following way: an observable O(x), depending on the variable x,  is scale invariant under the arbitrary change x $\to~\lambda$x, if there is a number $\mu(\lambda)$ such that O(x) = $\mu$O($\lambda$x). $\lambda$ is the fundamental scaling ratio. The solution of  O(x) is the power law:
\be
 O(x) = C  x ^ { \alpha}
\ee
 where $\alpha$ = -ln$\mu$/ln$\lambda$.
 
The relative value of the observable, at two different scales, depends only on $\mu$, the ratio of the two scales O(x)/O($\lambda$x) and does not depend on x. We have therefore " a continuous translational invariance expressed on the logarithms of the variables".
If the distribution of the logarithm of O(x) versus the logarithm of x, displays a straight line, we expect that a relation exists allowing the fit by the DSI model, and allowing the calculation of the masses, starting from the lowest mass.
\subsection{The discrete scale invariance characteristic of hadronic masses}
 The {\it discrete} scale invariance (DSI), in opposition to the 
 {\it continuous} scale invariance, is observed when the scale invariance is only observed for specific choices of $\lambda$. 
Its signature is the presence of power laws with complex exponents $\alpha$ inducing log-periodic corrections to scaling \cite{sornette}. In case of DSI, the $\alpha$ exponent is \\
\be
\hspace*{8.mm}\alpha = -ln\mu/ln~\lambda + i 2n\pi/ln\lambda \\
\ee
where n is an arbitrary integer. The {\it continuous} scale invariance is obtained for the special case n = 0, then $\alpha$ becomes real.

The critical exponent "s" is defined by $\mu$ = $\lambda^{s}$. Defining $\Omega$ = 1/ln$\lambda$, we obtain $\alpha$ = -s + i 2$\pi \Omega$. 

The most general form of distributions following DSI was given by Sornette \cite{sornette}. We apply it to the ratio of two hadronic adjacent masses of a given species, "f(r)", where "r" is the rank of the distribution.
Following the most general form of the mass ratio distribution f(r):\\
\be
 f(r) = C\hspace*{1.mm}(|r-r_{c}|)^{s}\hspace*{1.mm}[1 + a_{1}\hspace*{1.mm}cos(2\pi\hspace*{1.mm}\Omega\hspace*{1.mm}
ln\hspace*{1.mm}(|r-r_{c}|) + \Psi)]\\
\ee
where we have omitted the imaginary part of f(r). 
 
 C is a normalization constant. $a_{1}$ measures the amplitude of the log-periodic correction to continuous scaling, and $\Psi$ is a phase in the cosine.
"$r_{c}$" is the  critical rank, which describes the transition from one phase to another. It is underdetermined, but widely larger than the experimental  "r" values. This is the general situation of hadronic mass ratios.
Such undetermination has very few consequences on all parameters, except
on $\Omega$. When increasing "$r_{c}$" from 30 to 40, $\Omega$ increases approximately by a factor 1.36. Therefore "$r_{c}$"  was arbitrarily fixed to 
"$r_{c}$" = 40 for all following studies, the same value as the one used in the application of fractals for the study of hadronic masses \cite{frachadron}.

In summary, the signature of scale invariances is the existence of power laws. The exponent $\alpha$ is real if we have continuous scale invariances  and is complex in case of discrete scale invariances, and then gives rise to log-periodic corrections.

 We first plot the log of the studied quantity, versus the log of its rank (ln(R)), defined as being the mass number from the lighter to the heavier studied mass. All masses are expressed in MeV.
 
Then we study the ratio of successive values fitted by equation (3).  
\section{Application to nuclei masses}
In all following applications, when the fundamental nuclei masses are studied, they are introduced with increasing  values of the variable Z (number of protons),  N (number of neutrons) , or the mass number A.  The masses are taken from the Atomic Mass Table \cite{wapstra}. They are very precise, therefore the errors, which are introduced in the computations, are not visible.
\subsection{Application to light nuclei with Z (then N) =1, 2, or 3}
\begin{figure}[h]
\caption{Log-log distribution of light nuclei masses. Full triangles (green on line) show the distribution for Z = 1 nuclei, A increasing from 1 up to 7. Full circles (red on line) show the distribution for Z = 2 nuclei, A increasing from 3 up to 10. Full squares (blue on line) show the distribution for Z = 3 nuclei, A increasing from 3 up to 12, (see text). } 
\hspace*{-3.mm}
\scalebox{1}[1.5]{
\includegraphics[bb=23 234 525 550,clip,width=0.45\textwidth]{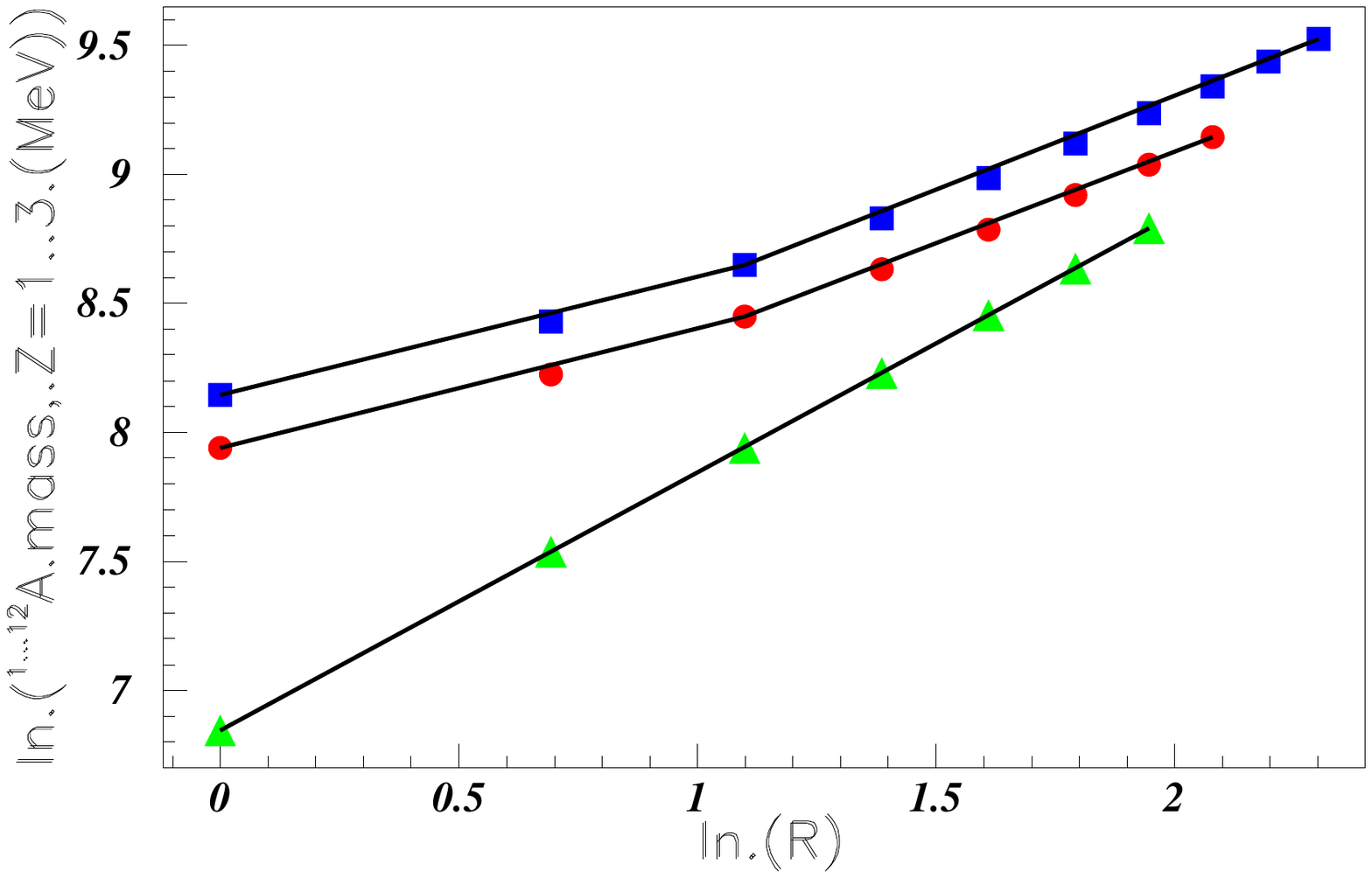}}
\end{figure}
\begin{figure}[ht]
\caption{Ratios of $m_{n+1}/m_{n}$ masses of several light nuclei. Full triangles (green on line) show the distribution of Z = 1 nuclei, A increasing from 1 up to 7. Full circles (red on line) show the distribution of Z = 2 nuclei,  A increasing from 3 up to 10. Full squares (blue on line) show the distribution of Z = 3 nuclei, A increasing from 3 up to 12. } 
\hspace*{-3.mm}
\scalebox{1}[1.5]{
\includegraphics[bb=12 234 525 550,clip,width=0.45\textwidth]{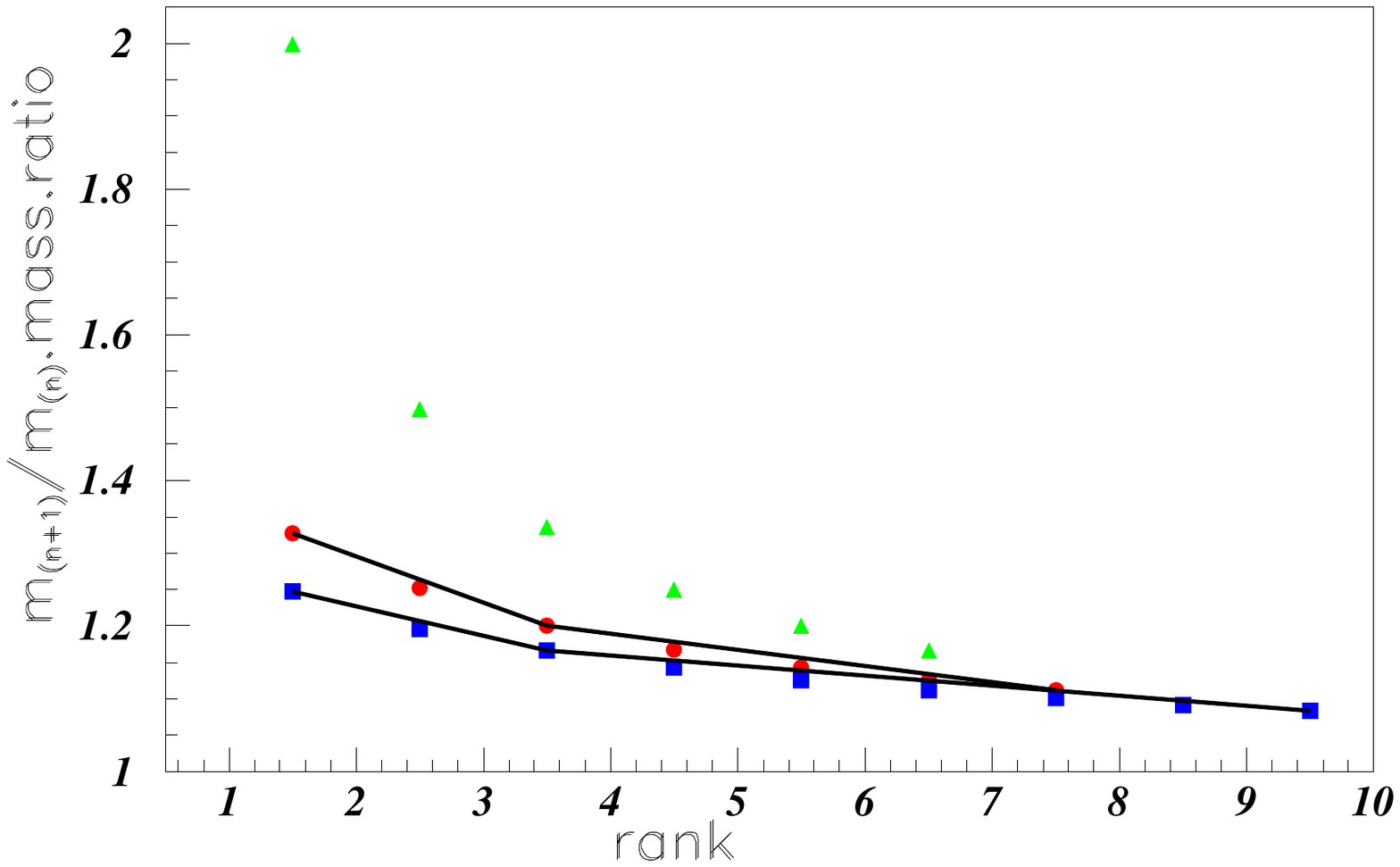}}
\end{figure}
Figure 1 shows the log-log distribution of light nuclei masses \cite{wapstra} for constant number of protons (Z=1, 2, or 3) and increasing variable number of neutrons. Full triangles (green on line) show the distribution of Z~=~1 nuclei, A increasing from 1 (R~=~1) up to 7 (R~=~7). Full circles (red on line) show the distribution of Z~=~2 nuclei, A increasing from 3 (R~=~1) up to 10 (R~=~8). Full squares (blue on line) show the distribution of Z~=~3 nuclei, A increasing from 3 (R~=~1) up to 12 (R~=~10). These last data, full squares (blue on line), are shifted vertically by the quantity log(shift(MeV)) = 0.2, in order to separate these data from the Z~=~2 data. The distributions exhibit straight lines, which indicate fractal properties.

The nice alignements of the log-log distributions suggest to extrapolate them in order to predict the masses of the next nuclei. The resulting results are given in table~1, where all masses are in MeV. The third column indicates the last (heavier) known experimental mass, followed by the same mass calculated using the previous masses and the linearity of the log-log distributions. The relative precision between both is given in the next column, and finally the predicted mass of the still unobserved nucleus is given in the last column.

The shape of the  distributions, however, shown in figure~2, does not correspond to the oscillatory variation described by equation (3).  The assumption could be advanced, that the low value of the maximum rank, due to the small number of such nuclei, could be responsible for the oscillation deficiency. This is probably not the case, since much lighter particles as quarks and leptons, display clear oscillations, in spite of a smaller number of elements, therefore a smaller maximum rank \cite{arxiv1104}. All data describing the
$m_{n+1}/m_{n}$ mass ratio, will be plotted at rank n+1/2. 
\begin{table}[t!]
\begin{center}
\caption{Experimental and "calculated" last observed masses with Z = 1, 2, and 3 nuclei corresponding to figure 1, then N = 1, 2, and 3 nuclei corresponding to figure 3. Each line ends by the next predicted, but (still) not observed mass. All masses are in MeV; the empty markers correspond to full markers of figure 1 and 3.}
\label{Table 1}
\begin{tabular}[t]{c c c c c c c c}
\hline
      &           & last known& &mass&next predicted\\
Fig.&Marker&exper.& calcul&rel gap.&mass\\
\hline
1&Z=1 \textcolor{green}{$\bigtriangleup$}&6569.0&6568.63&5.3 10$^{-5}$&$^{8}$H 7507.4\\
 &\hspace*{3.2mm}2 \textcolor{red}{$\bigcirc$}&9362.6&9244.4& 1.3\%&$^{11}$He 10179\\
 &\hspace*{2.8mm}3 \textcolor{blue}{$\Box$}&11226.3&11095.2&1.2\%&$^{13}$Li 12034\\
\hline
3 &N=1 \textcolor{green}{$\bigtriangleup$}&5629.9&5631.4&2.6 10$^{-4}$&$^{7}$C 6567.5\\
&\hspace*{3.2mm}2 \textcolor{red}{$\bigcirc$}&7483.8&7385.4&1.3\%&$^{9}$N 8312.3\\
&\hspace*{3.1mm}3 \textcolor{blue}{$\Box$}&9350.0&9204.4&1.6\%&$^{11}$O 10136\\
\hline
\end{tabular}
\end{center}
\end{table}

Figure~3 shows the log-log distribution for light nuclei, constant number of neutrons "N", and variable number of protons.  Full triangles (green on line) show the distribution of N = 1 nuclei, A increasing from 1 (R~=~1) up to 6 (R~=~6). Full circles (red on line) show the distribution of N~=~2 nuclei, A increasing from 3 (R~=~1) up to 8 (R~=~6). Full squares (blue on line) show the distribution of N~=~3 nuclei, A increasing from 4 (R~=~1) up to 10 
(R~=~7).  Here again straight lines indicate nice fractal properties. The shape of the three distributions, is like the one drawn for constant "Z" and shown in figure~ 1. 
\begin{figure}[ht]
\caption{Log-log distribution of light nuclei masses for constant number of neutrons "N". Full triangles (green on line) show the distribution of N~=~1 nuclei, A increasing from 1 up to 6. Full circles (red on line) show the distribution of N~=~2 nuclei, A increasing from 3 up to 8. Full squares (blue on line) show the distribution of N~=~3 nuclei, A increasing from 4 up to 10.} 
\hspace*{-3.mm}
\scalebox{1}[1.5]{
\includegraphics[bb=23 234 525 550,clip,width=0.45\textwidth]{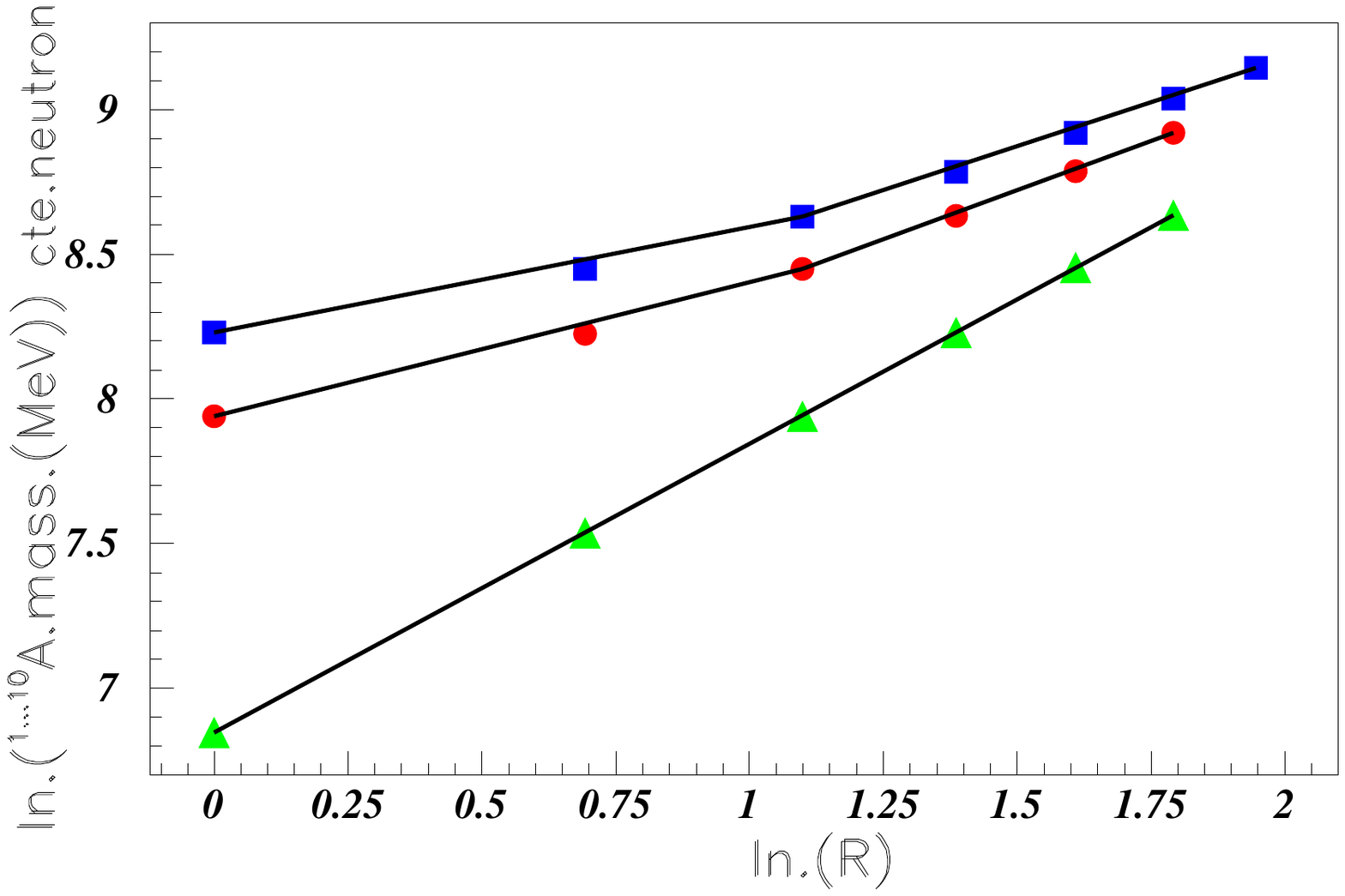}}
\end{figure}

Here again, the nice alignements of the log-log distributions suggest to extrapolate them in order to predict the masses of next nuclei. The resulting results are given in table~1, where all masses are in MeV. The third column indicates the last experimental mass, followed by the corresponding calculated value.  The calculated masses use again the linearity of the previous masses in the log-log distributions. The relative precision between both is given in the next column, and finally the predicted masses of the (still) unobserved nuclei are given in the last column.
\subsection{Application to  A = 10 nuclei masses}
\begin{figure}[ht]
\caption{Log of A = 10  nuclei masses versus the log of the rank "R".}
\hspace*{-3.mm}
\scalebox{1}[1.5]{
\includegraphics[bb=9 240 520 550,clip,width=0.45\textwidth]{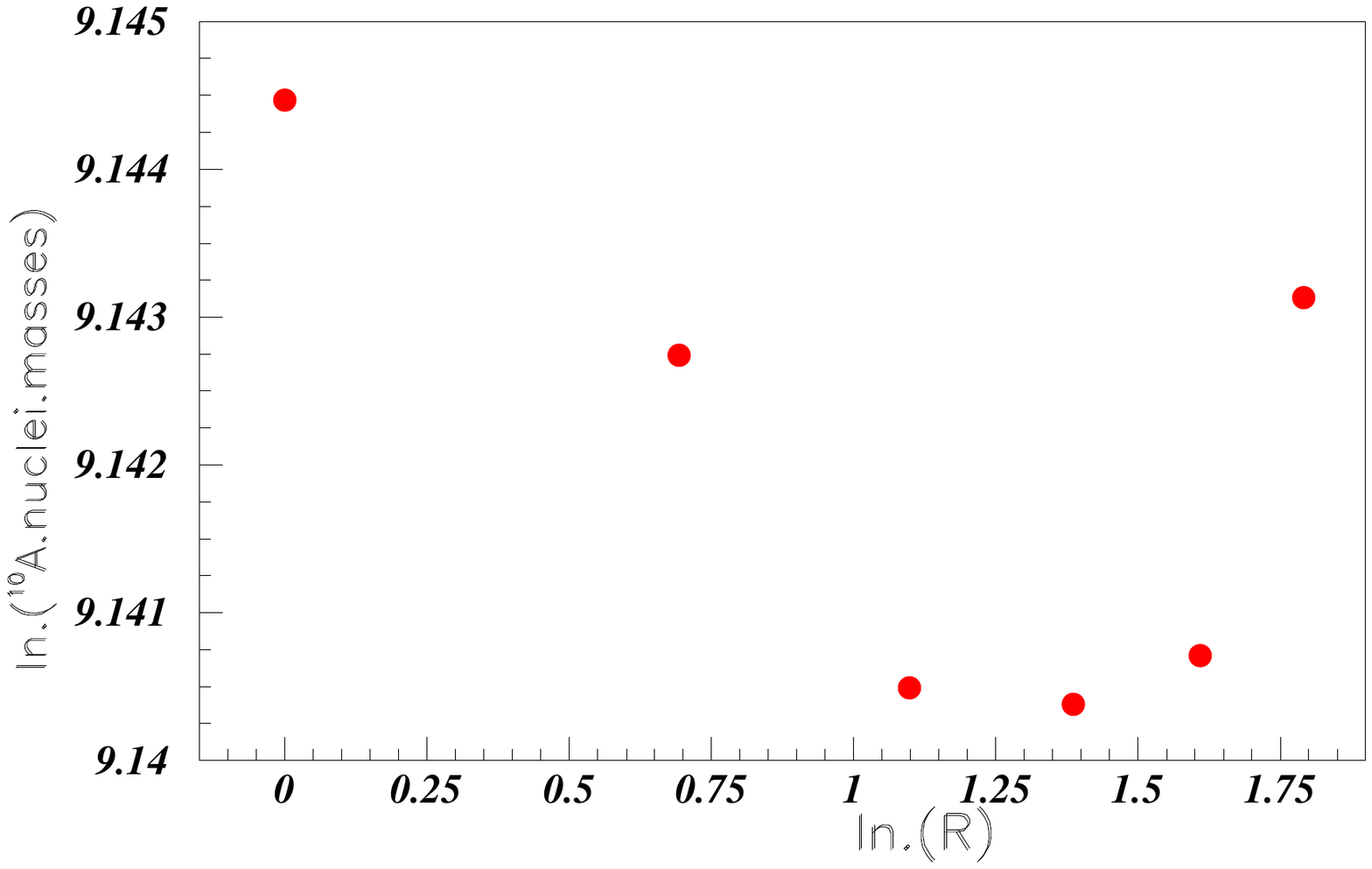}}
\end{figure}
Only six masses of ten nucleons are reported in the PDG table \cite{wapstra}. Figure~4 shows that the data in the log-log plot are not aligned, therefore these data should not follow the fractal properties.   We observe however that the  $m_{n+1}/m_{n}$ mass ratio of A~=~10 nuclei, versus the rank R, shown in figure~5, is rather well fitted by equation (3). We conclude that the
$m_{n+1}/m_{n}$ mass ratio distribution, as described by equation (3), is more general, and could eventually be not restricted to data following fractal properties.
\begin{figure}[ht]
\caption{ Ratios of $m_{n+1}/m_{n}$ masses versus the rank, for A = 10 nuclei. } 
\hspace*{-3.mm}
\scalebox{1}[1.5]{
\includegraphics[bb=1 230 520 550,clip,width=0.45\textwidth]{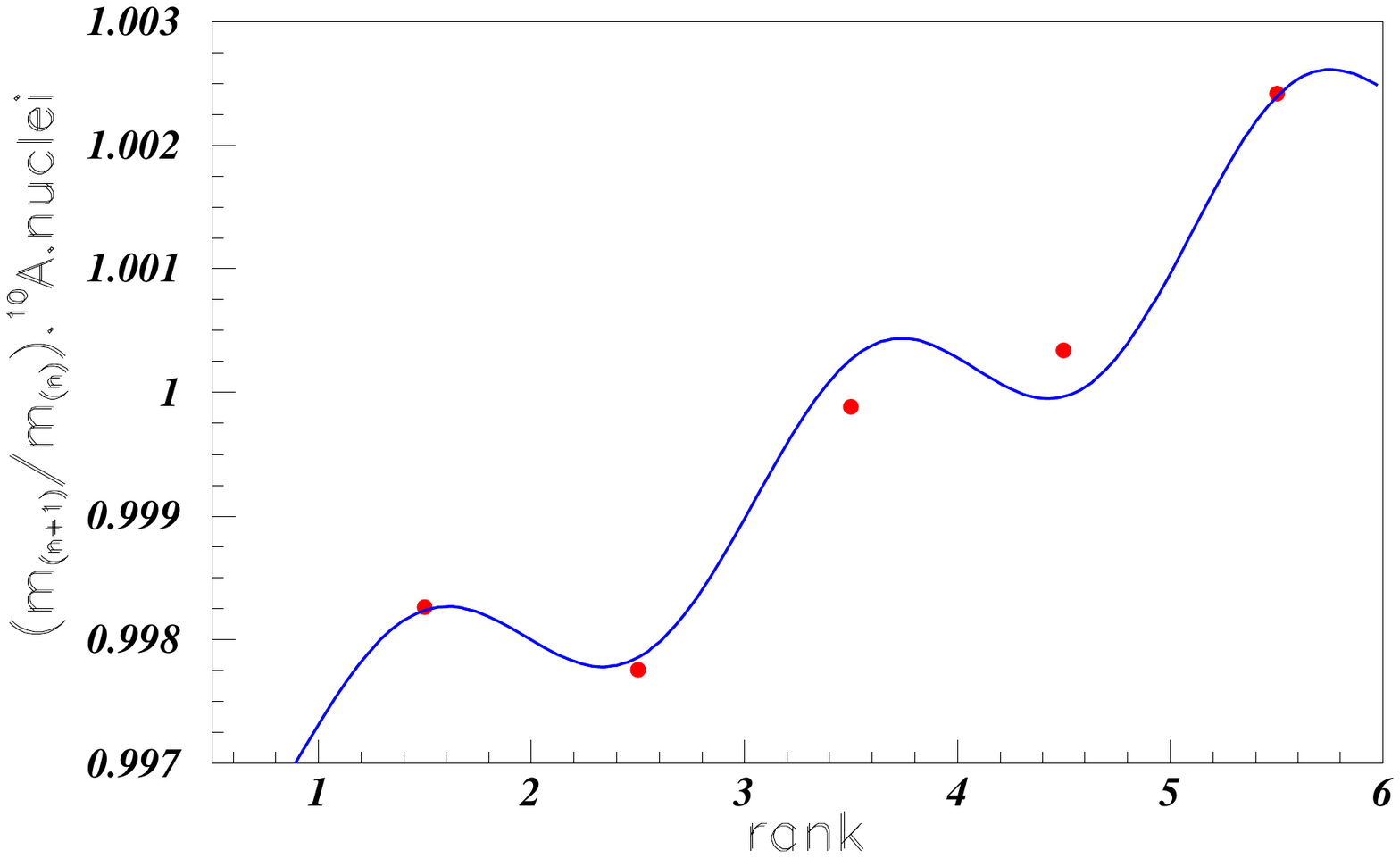}}
\end{figure} 
\subsection{Application to A = 21 nuclei masses}
\begin{figure}[ht]
\caption{Log of A = 21 nuclei masses versus the log of the rank "R".}
\hspace*{-3.mm}
\scalebox{1}[1.5]{
\includegraphics[bb=4 240 514 547,clip,width=0.45\textwidth]{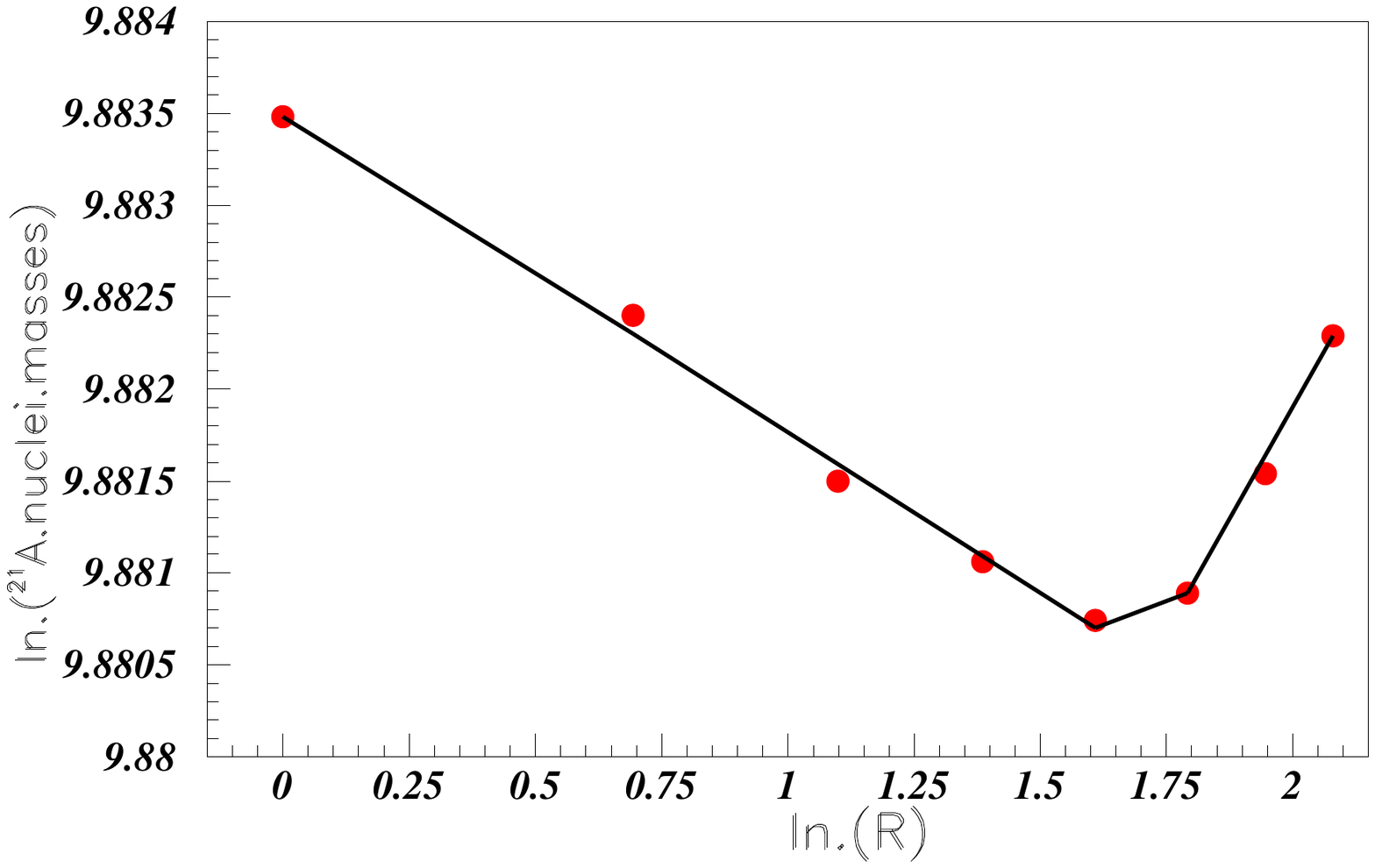}}
\end{figure}
Figure~6 shows three log-log distributions for A~=~21 nuclei masses. In this figure, the first five masses, and the last  three masses are aligned, which shows fractal properties.
The corresponding  $m_{n+1}/m_{n}$ mass ratio versus the rank "R", is shown in figure~7. All seven data points (except one (the fourth)), are well described by the fit. Although the amplitude of the oscillations, given by the parameter "a$_{1}$" of equation (3), is poorly determined, the parameters "$\Omega$" and "s", therefore the important parameters "$\alpha$" and "$\lambda$" are well determined by the data.
\begin{figure}[ht]
\caption{Ratios of $m_{n+1}/m_{n}$ masses, versus the rank "R", for A~=~21 nuclei.}
\hspace*{-3.mm}
\scalebox{1}[1.5]{
\includegraphics[bb=8 236 520 540,clip,width=0.45\textwidth]{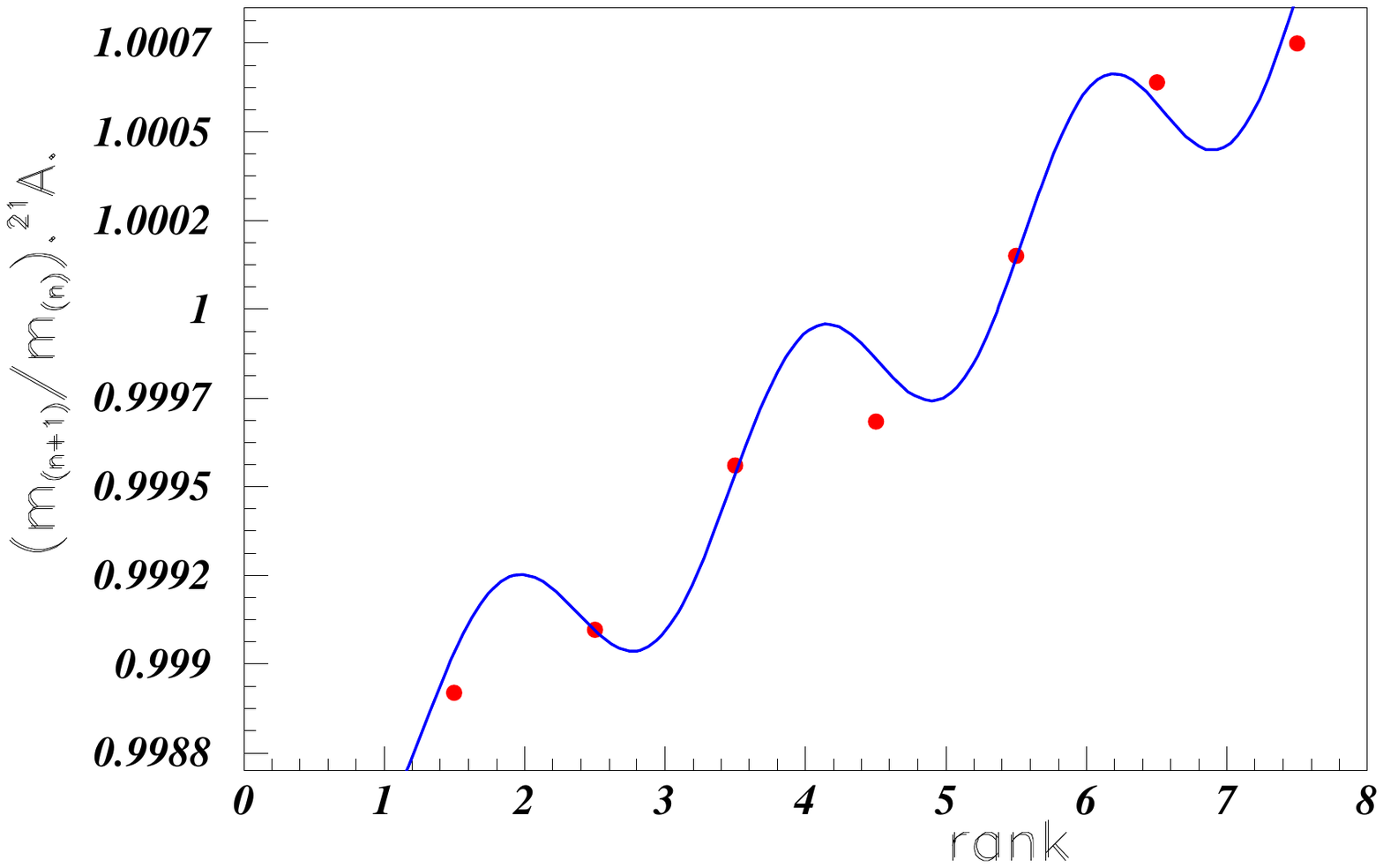}}
\end{figure}
\subsection{Application to nuclei with masses A = 38, 40, and 42}
\begin{figure}[ht]
\caption{Log-log distribution of three nuclei masses for increasing Z values. Full circles (red on line) show data for A=40. Full squares (blue on line) show results for A = 42, the log values of these data are shifted by -0.048 in order to enter in the same figure. Full triangles (green on line) show results for A = 38; the log values of these last data are shifted by +0.0505 in order to enter in the same figure. } 
\hspace*{-3.mm}
\scalebox{1}[1.5]{
\includegraphics[bb=1 245 520 545,clip,width=0.45\textwidth]{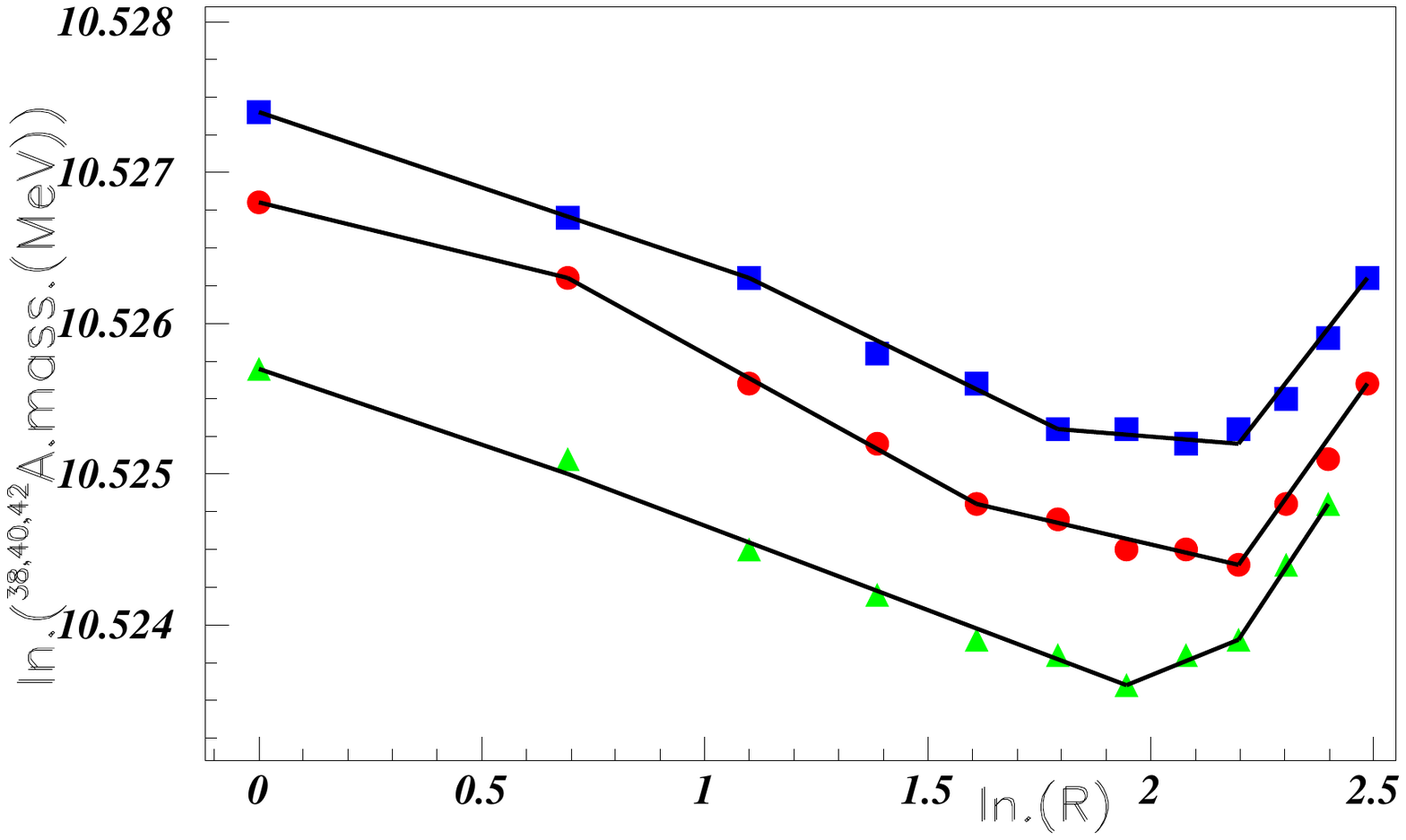}}
\end{figure}
Figure 8 shows the log-log distribution of the
 A~=~38, 40, and 42 nuclei masses (for increasing Z values), calculated with help of the mass excess table \cite{wapstra}. The Z values increase from 12 up to 22 for A~=~38 (rank~=1 up to 11), increase from 12 up to 23 for A~=~40 (rank~=~1 up to 12), and increase from 13 up to 24 for A~=~42 (rank~=~1 up to 12). Several alignements manifest the fractal properties of these data. In order to clarify the figure, the log values of these data for A = 38 and 42 are shifted respectively by +0.0505 and -0.048 in the ordinate scale.
\begin{figure}[ht]
\caption{ Ratios of $m_{n+1}/m_{n}$ masses versus the rank, for nuclei close to A~=~40. Full circles (red on line) show data for A~=~40 and increasing Z values; full squares (blue on line) show results for A = 42; full triangles (green on line) show results for A = 38.} 
\hspace*{-3.mm}
\scalebox{1}[1.5]{
\includegraphics[bb=1 210 523 550,clip,width=0.45\textwidth]{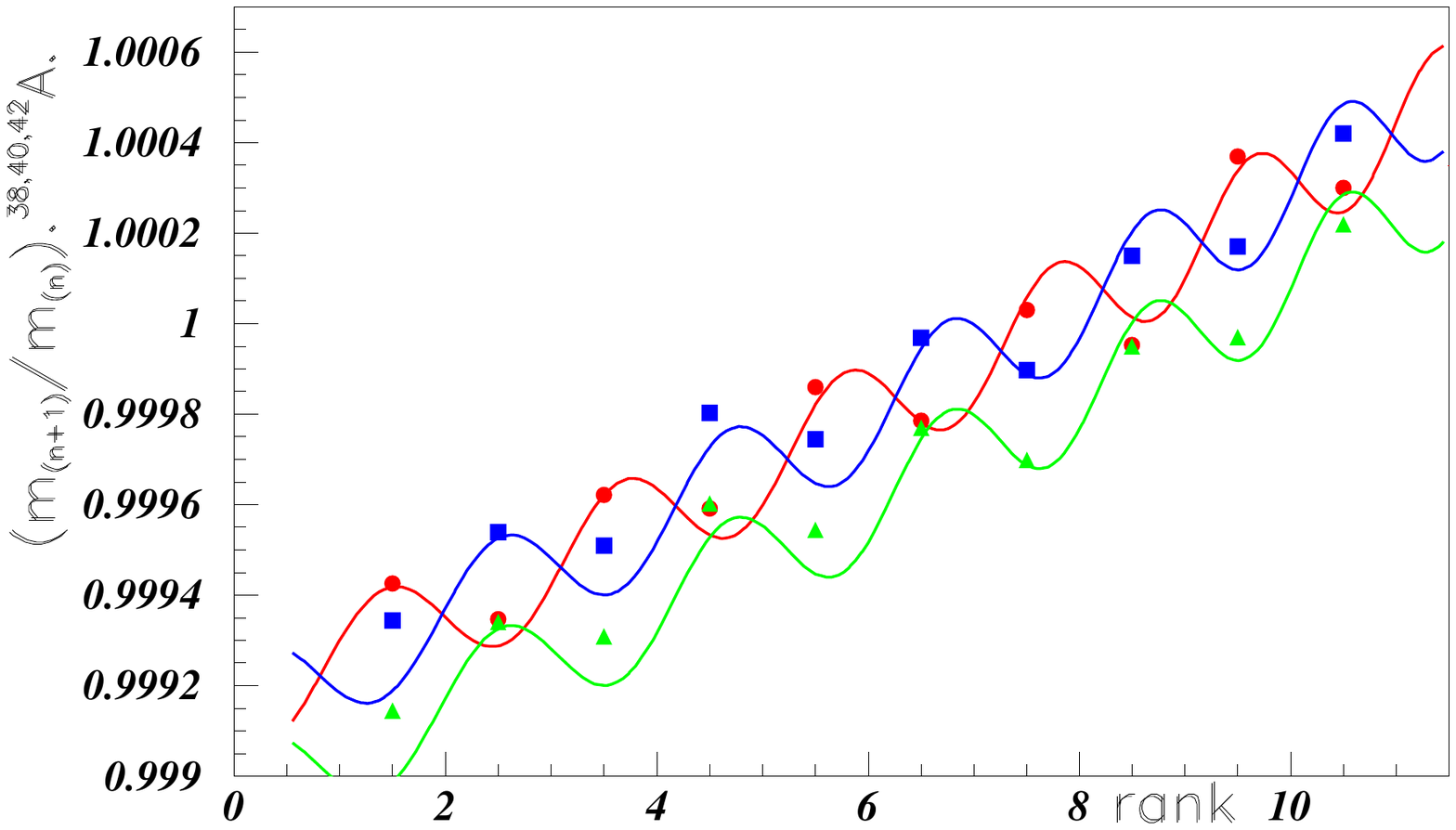}}
\end{figure}

Figure~9 shows the $m_{n+1}/m_{n}$ mass ratios, versus the rank, for nuclei in the A$\approx$40 nuclei and increasing Z values. Full circles (red on line) show data for A=40; full squares (blue on line) show results for A = 42; full triangles (green on line) show results for A = 38. The experimental ratio values of these last data, and the calculated curve, are renormalized by 0.9998 in order to be distinguished from A = 42 nuclei data. The fit gives rise to clearly observed nice oscillations, which describe the data in the whole experimental range.  All three curves are obtained using equation (3) and same parameters, except the phase $\Psi$, different for $^{40}$A and equal for $^{38}$A and $^{42}$A.  As a matter of fact, the A~=~38 and 42 distributions are in phase in figure 9, when the A~=~40 distribution is in an opposite phase. 

\begin{figure}[ht]
\caption{Log of some nuclei masses around $^{40}$Ca versus the log of the rank "R". Insert (a) shows the distribution for calcium isotopes (full circles, red on line), the distribution of sulfur isotopes (full squares, blue on line), and the distribution of 
N~=~20 isotones (full triangles, green on line), slightly shifted (see text); insert (b) shows  the distribution for A~=~40 nuclei.}
\hspace*{-3.mm}
\scalebox{1}[1.5]{
\includegraphics[bb=5 123 525 553,clip,width=0.45\textwidth]{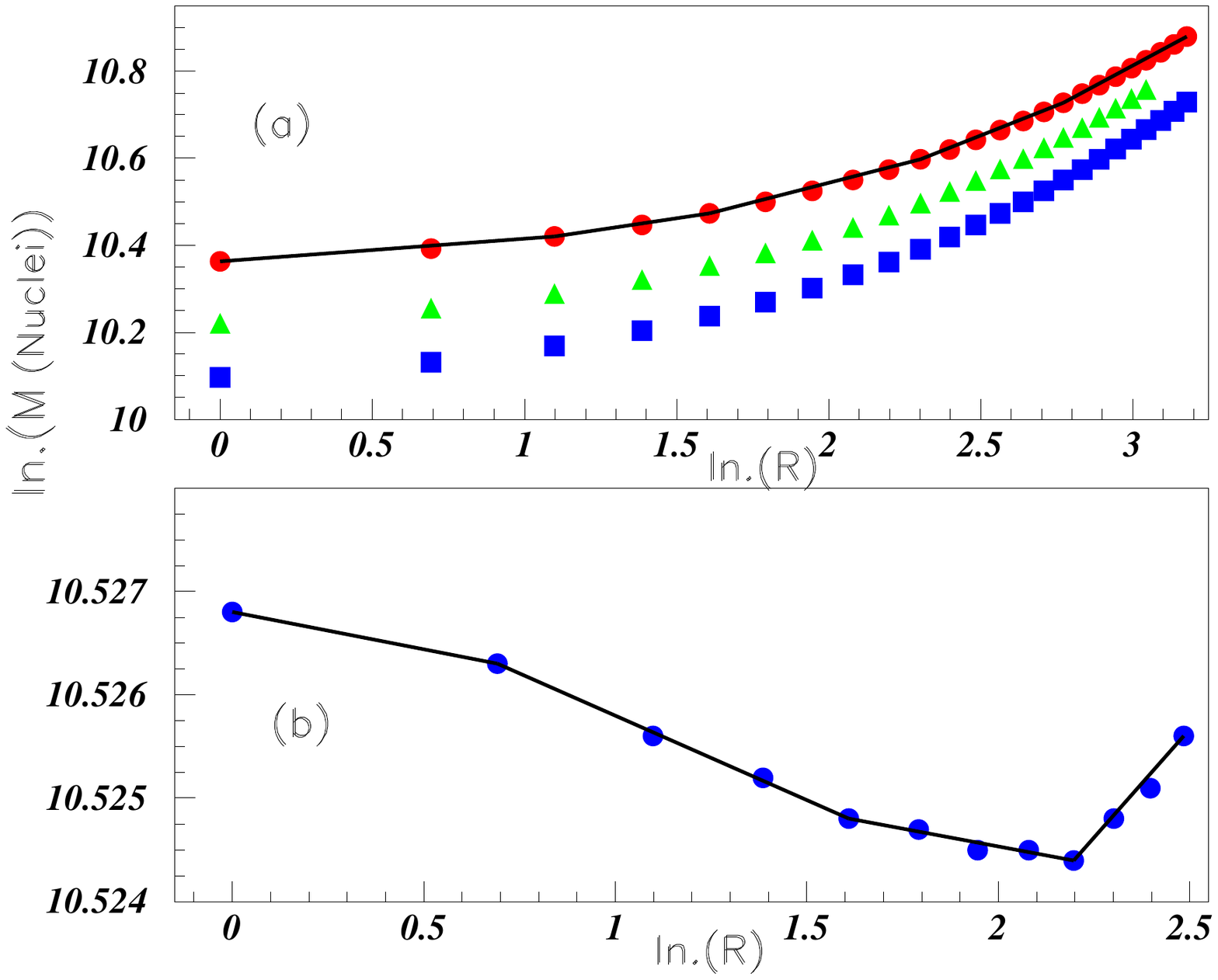}}
\end{figure}
Figure 10 shows several log-log distributions for constant Z nuclei around $^{40}$Ca.  Figure~10(a) shows the distribution for calcium isotopes (full circles, red on line), the distribution of sulfur isotopes (full squares, blue on line), and the distribution of N~=~20 isotones (full triangles, green on line). These last data are shifted by +0.05  in order to separate them from the sulfur data. Figure~10(b) shows  the distribution for A = 40 nuclei, already shown in figure~8 where the variation displays several straight lines for the distribution corresponding to the mass variation for A = 40 nuclei. The fractal property of A~=~40 nuclei is more effective, than it was for calcium and sulfur isotopes, and N~=~20 isotones; it is however less effective than it was for light nuclei.

The linearity at the end of the distributions shown in figure~8, allows us to tentatively extrapolate the masses and predict the masses of still unobserved nuclei.
Table~II gives the last experimental observed masses of the A = 38, 40, and 42 nuclei, compared to the  "calculated" masses. The calculated masses use the linearity of the previous masses in the log-log distribution. The relative errors between both lie between 1.10$^{-4}$ and  2.10$^{-4}$. The last column shows the predicted mass of the next, still unobserved nuclei.
\begin{table}[h]
\begin{center}
\caption{Experimental and "calculated" last observed masses for $^{38,40,42}$A nuclei corresponding to figure~8, followed by the next predicted, but still not observed, mass. All masses are in MeV. The empty markers here, correspond to the full markers in the figures.}
\label{Table 2}
\begin{tabular}[t]{c c c c c c}
\hline
      & last &exprim. &mass&next mass\\
Marker&exper.& calcul&rel gap.&predicted\\
\hline
A=38 \textcolor{green} {$\bigtriangleup$} & 35394.0 & 35397.6 & 1.0 10$^{-4}$
&  $^{38}$ V 35405\\
 \hspace*{4.2mm}40  \textcolor{red} {$\bigcirc$} & 37257.9 & 37249.7 &  2.1 10$^{-4}$ &                $^{40}$Cr 37268\\
 \hspace*{2.8mm}42 \textcolor{blue} {$\Box$} & 39115.8 & 39108.7 & 1.8 10$^{-4}$  &     $^{42}$Mg 39128\\
\hline
\end{tabular}
\end{center}
\end{table}
\subsection{Application to masses  of A = 99 and 100 nuclei}
\begin{figure}[ht]
\caption{Log of A = 100 (full circles, red on line) and A~=~99 nuclei masses (full squares, blue on line) versus the log of the rank "R".}
\hspace*{-3.mm}
\scalebox{1}[1.5]{
\includegraphics[bb=8 240  515 543,clip,width=0.45\textwidth]{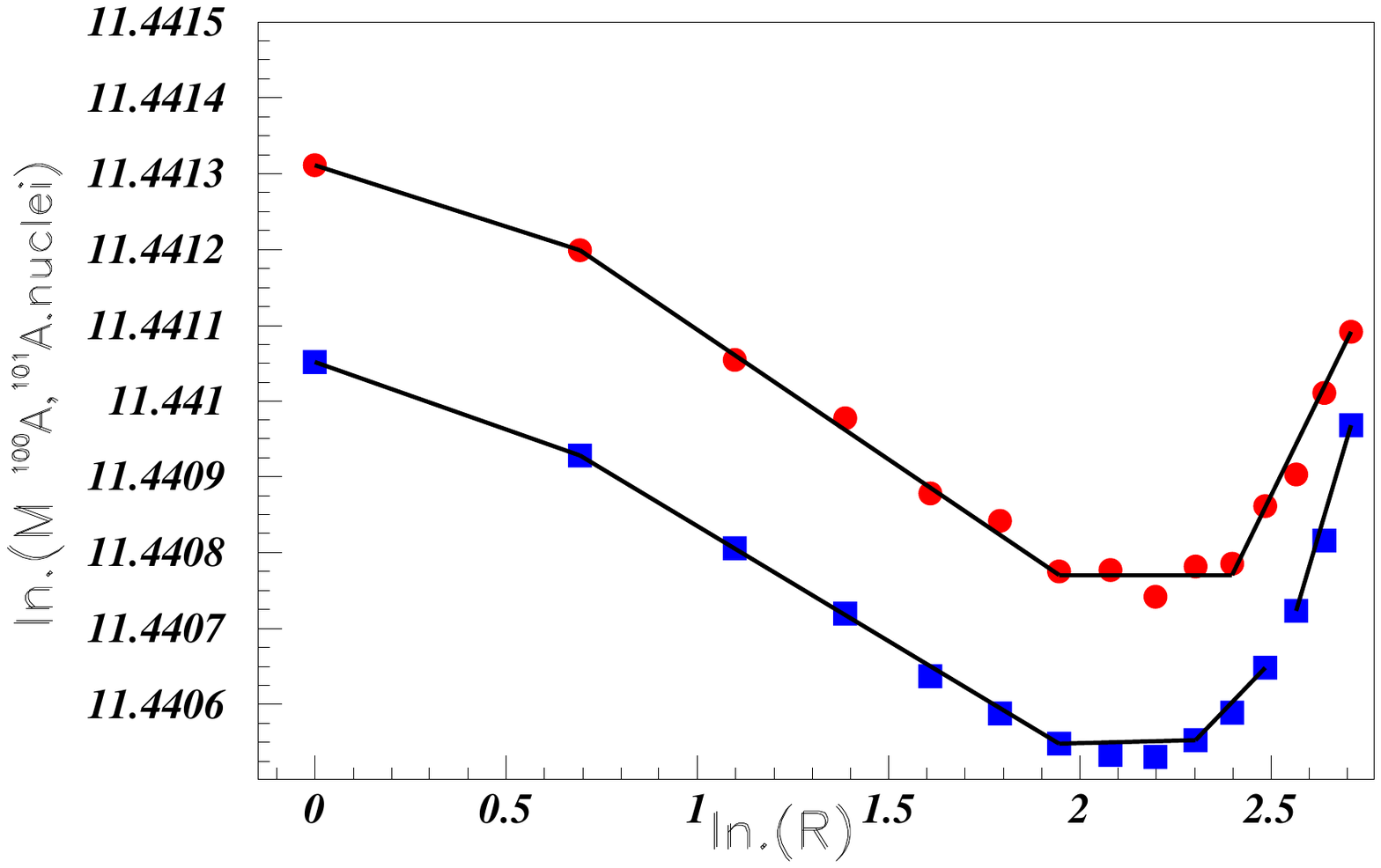}}
\end{figure}
Figure~11 shows that many masses in the log-log plot of A~=~100 nuclei (full circles, red on line) and in the log-log plot of A~=~99 nuclei (full squares, blue on line), can be described by a few straight lines. The A~=~99 distribution is shifted by 0.00978 in the y coordinate.
These data should therefore follow the fractal properties. 
Here Z increases, for both distributions, from 36 (rank 1) up to 50 (rank 15).
 The  $m_{n+1}/m_{n}$ mass ratio of A = 100 nuclei (full circles, red on line), are shown versus the rank R in figure~12. The  $m_{n+1}/m_{n}$ mass ratio of A = 99 nuclei (full squares, blue on line), are shown versus the rank R in the same figure. They both  exhibit many oscillations, very well fitted by equation~(3) up to rank 15 (last rank). The oscillations are larger in case of even nucleus (A~=~100), than for odd nucleus (A~=~99). The ratio between both a$_{1}$ factors, which describe the oscillation amplitude, is equal to 2.6. 
 
\begin{figure}[ht]
\caption{ Ratios of $m_{n+1}/m_{n}$ masses versus the rank, for A = 100 nuclei (full circles, red on line) and A~=~99 nuclei (full squares, blue on line).} 
\hspace*{-3.mm}
\scalebox{1}[1.5]{
\includegraphics[bb=1 210 523 550,clip,width=0.45\textwidth]{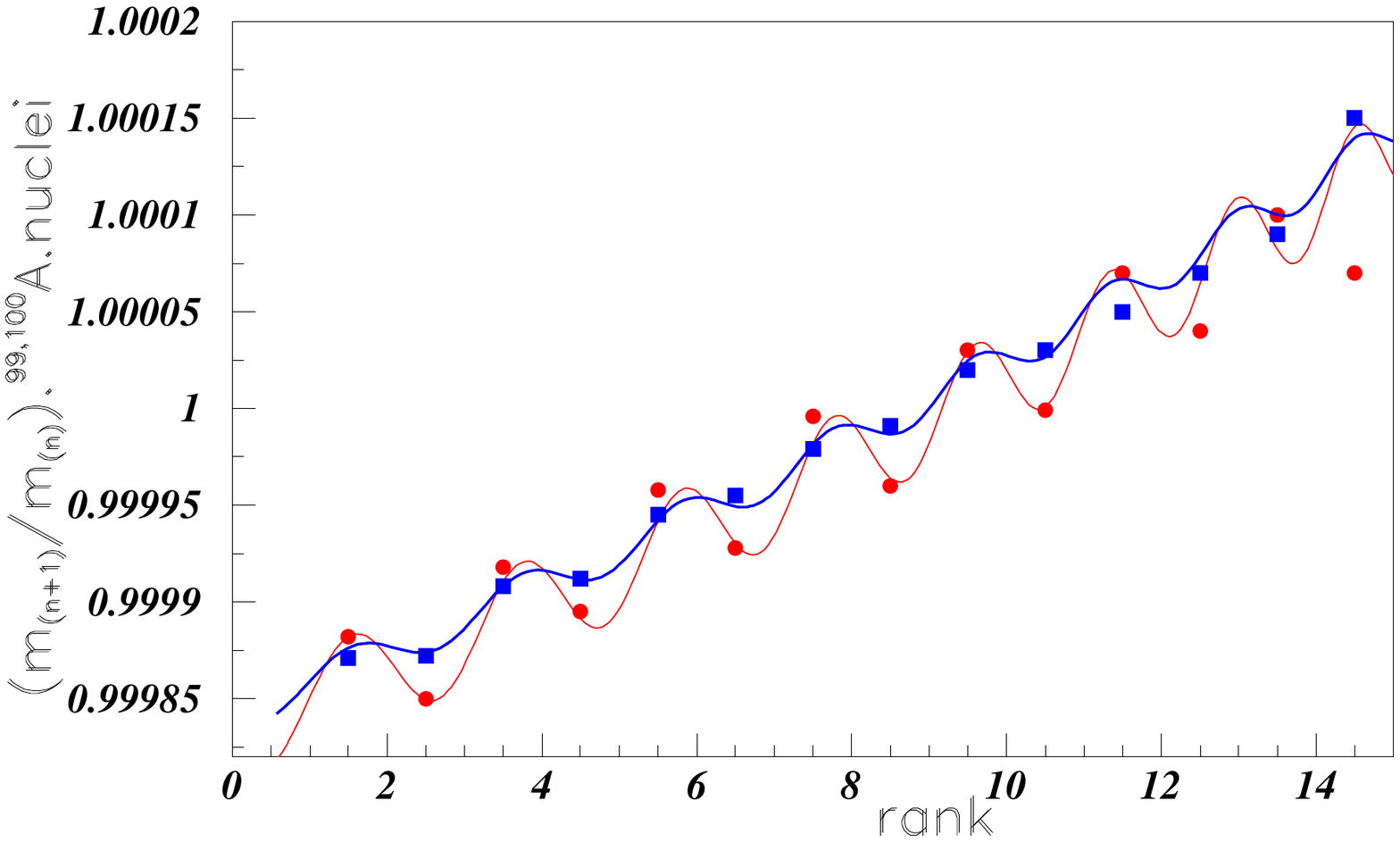}}
\end{figure}
\subsection{Application to some heavy nuclei with A = 158, 192, and 251}
 The next figure analyses the distributions for three heavier mass nuclei, in order to look at eventual fractal properties for intermediate and heavy mass nuclei. 
\begin{figure}[ht]
\caption{Log of some nuclei masses versus the log of the rank "R". Full circles (red on line), show the results for A = 158 nuclei, full squares (blue on line), show the results for A = 192 nuclei shifted by -0.195, and full triangles (green on line), show the results for A = 251 nuclei shifted by -0.4633.}
\hspace*{-3.mm}
\scalebox{1}[1.5]{
\includegraphics[bb=9 333 525 553,clip,width=0.45\textwidth]{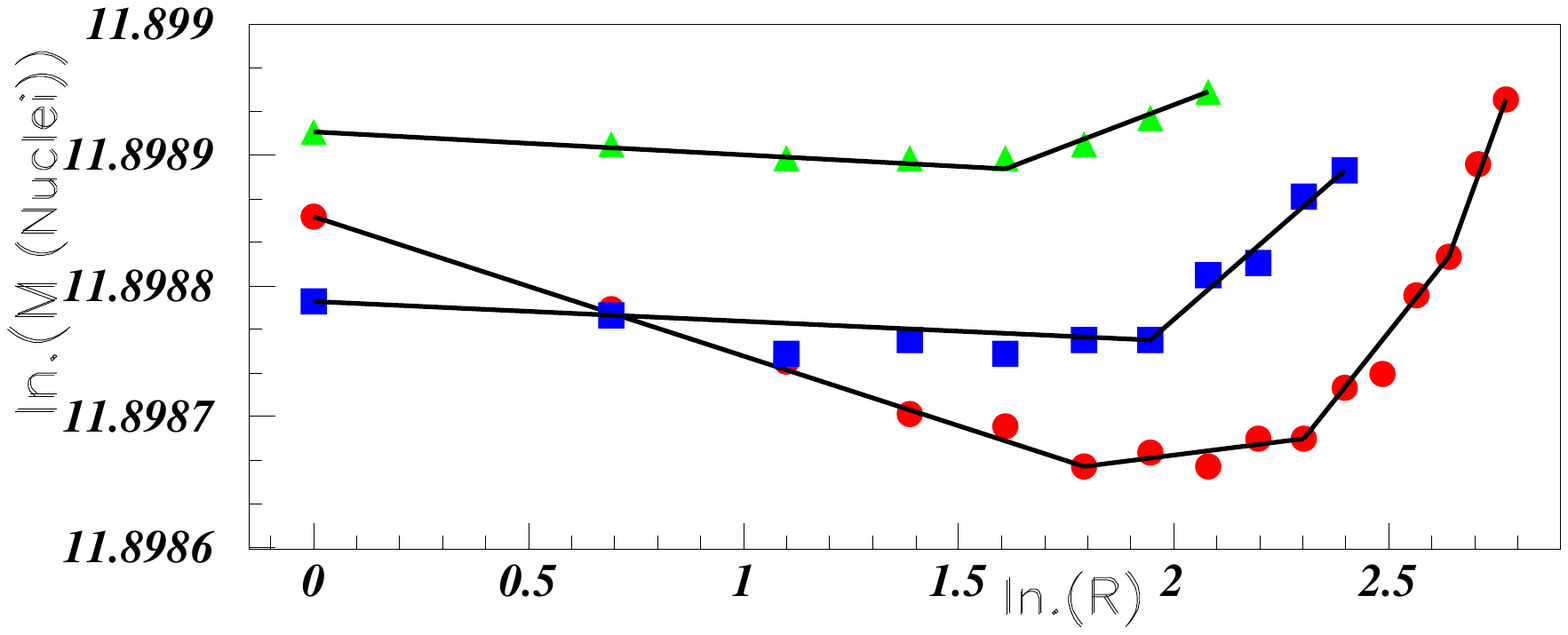}}
\end{figure}
Figure 13 shows three log-log distributions for increasing Z values:
full circles (red on line) show the distributions of A~=~158~ nuclei, Z increasing from 59 (rank 1) up to 74 (rank 16);
 full squares (blue on line) show the distribution of A = 192 nuclei, shifted by -0.195, Z increasing from 74 (rank 1) up to 84 (rank 11); and full triangles (green on line) show the distribution of A = 251 nuclei (green on line) shifted by -0.46353 in y coordinate, Z increasing from 96 (rank 1) up to 103 (rank 8). These translations are done for clarity, in order to accept all three distributions inside a same figure. We observe that distributions are well described by two straight lines for 
 A~=~251 and 192, and four straight lines for A~=~158 nuclei.
\begin{figure}[ht]
\caption{Ratios of $m_{n+1}/m_{n}$ masses for three different A values, versus the rank "R". Full circles (red on line), show the results for A = 158 nuclei with increasing Z values, full squares (blue on line), show the distribution for A = 192 nuclei,  and full triangles (green on line), show the distribution for A = 251 nuclei.}
\hspace*{-3.mm}
\scalebox{1}[1.5]{
\includegraphics[bb=0 225 525 550,clip,width=0.45\textwidth]{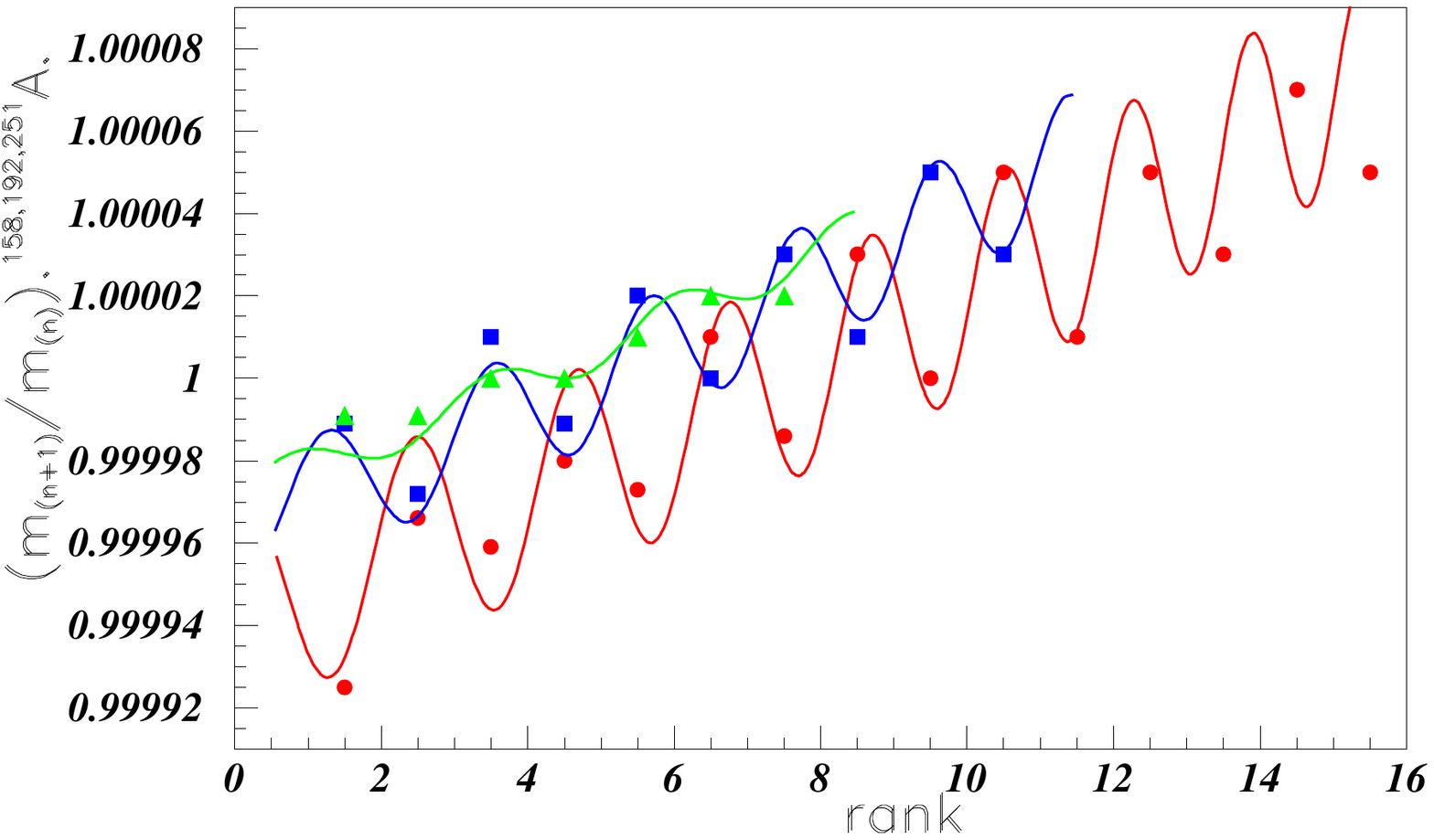}}
\end{figure}

Figure~14 shows the ratios of $m_{n+1}/m_{n}$ masses versus the rank "R", for the same three different A values, (and increasing Z values). Full circles (red on line), show the results for the A = 158 nuclei, full squares (blue on line), show the distriburion for A = 192 nuclei,  and full triangles (green on line), show the distribution for A = 251 nuclei. The data exhibit oscillations which reduces when the masses increase. Here again, the experimental data are well fitted  by equation (3).
\subsection{Application to masses  of A = 273 nuclei}
\begin{figure}[ht]
\caption{Log of A = 273 nuclei masses versus the log of rank "R".}
\hspace*{-3.mm}
\scalebox{1}[1.5]{
\includegraphics[bb=6 225 515 545,clip,width=0.45\textwidth]{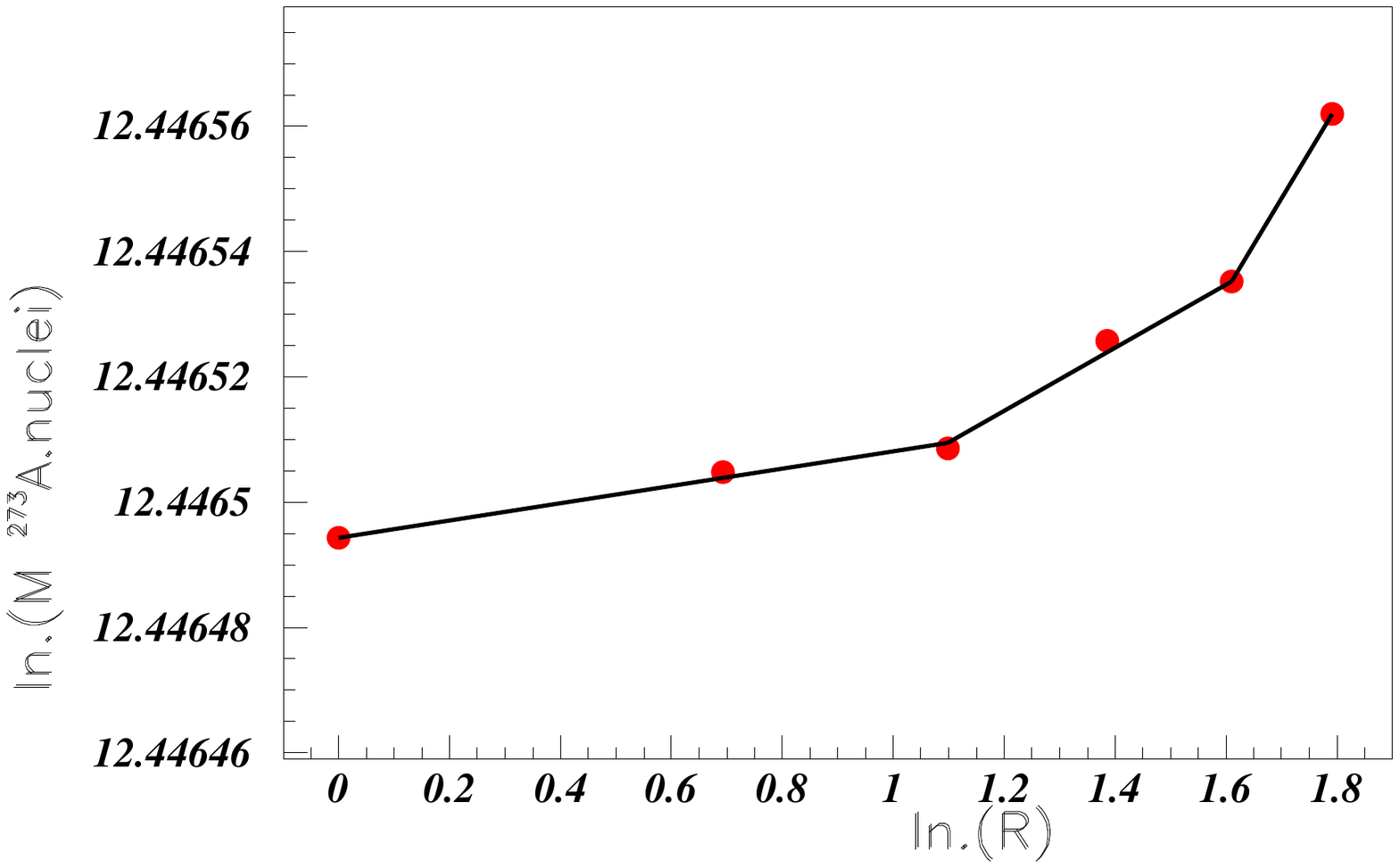}}
\end{figure}
\begin{figure}[ht]
\caption{ Ratios of $m_{n+1}/m_{n}$ masses versus the rank, for A = 273 nuclei. } 
\hspace*{-3.mm}
\scalebox{1}[1.5]{
\includegraphics[bb=1 210 523 550,clip,width=0.45\textwidth]{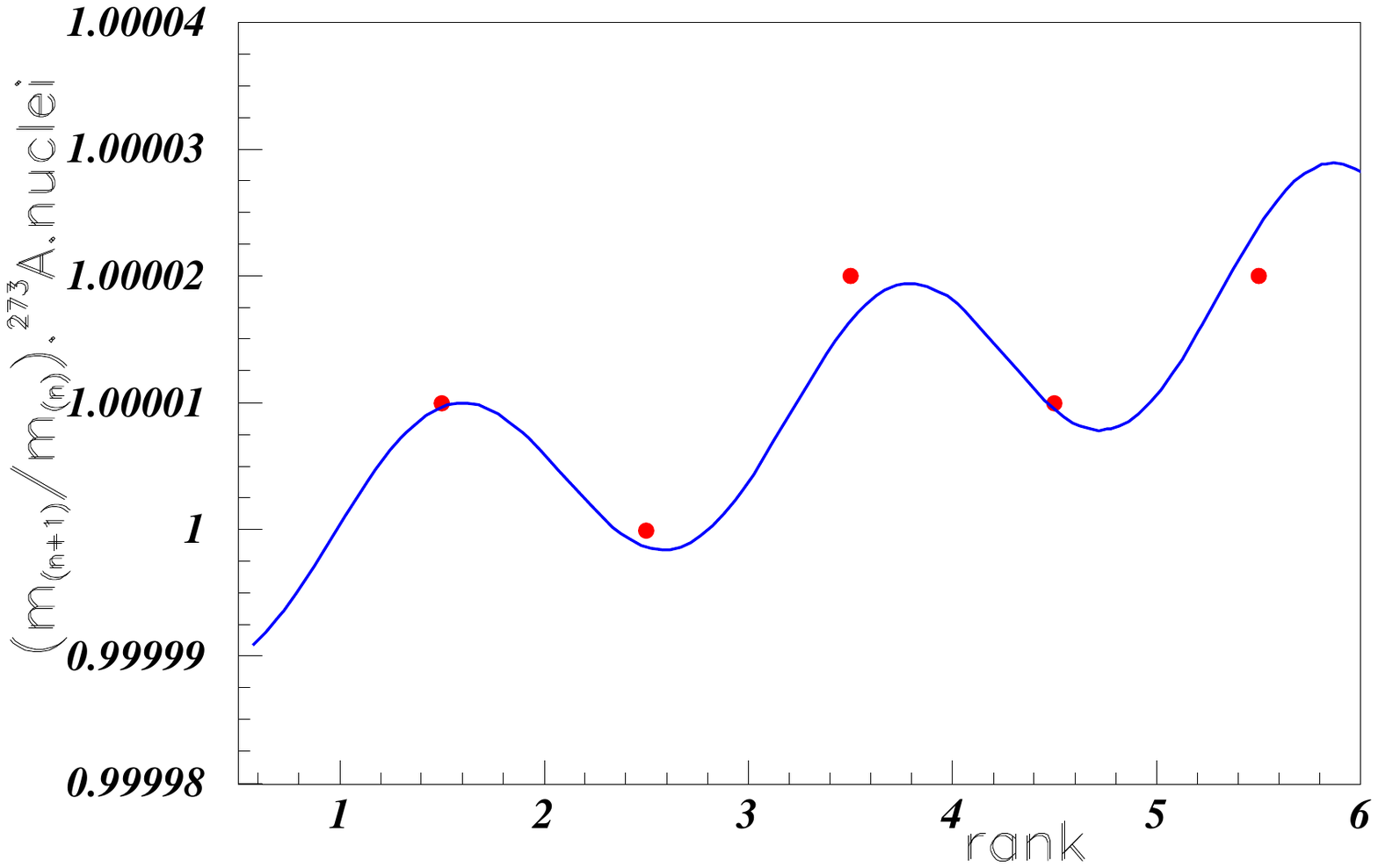}}
\end{figure} 
Only six masses of mass number A~=~273 nuclei are reported, and such small number does not allow a nice alignement of the log-log distribution shown in figure~15. The corresponding $m_{n+1}/m_{n}$ mass ratio versus  the rank, is shown un figure~16. As previously, the oscillations are well reproduced by the fit.\\

 A precise prediction, concerning the eventual superheavy and stable nuclei, is not possible. Indeed the masses of unstable heavy nuclei, in the region where such superheavy nuclei are expected to exist, can be calculated using the mass excess table \cite{wapstra} and the alpha particle energies of the desexcitation chains. The energies of the $\alpha$ particles of the desexcitation chains are  known \cite{oganessian}; they differ one another by only a few MeV (close to 10~MeV). Starting from A=289, Z=115 nucleus, we could get heavier masses (adding two more $\alpha$),
but not nuclear masses with N~=~184, and A~=~298, or 304, or 310. In the same way, starting from A~=~290, Z~=~115 will not allow to get the masses indicated above. 
\section{Application to excited level nuclei masses}
The excited level masses studied are introduced with increasing masses, starting from the fundamental one. The energy levels for different nuclei, are taken from several papers from F. Ajzenberg-Selove and T. Lauristen.
\subsection{Application to excited level energies for nuclei masses A=11,12, and 13}
 The excited level energies of A~=~11, 12, and 13 nuclei, are taken from the Ajzenberg-Selove table \cite{as111213}.
\begin{figure}[ht]
\caption{Log-log plot of excited level masses of $^{11}$C in full circles (red on line) and  $^{11}$B in full squares (blue on line) .} 
\hspace*{-3.mm}
\scalebox{1}[1.5]{
\includegraphics[bb=6 230 530 550,clip,width=0.45\textwidth]{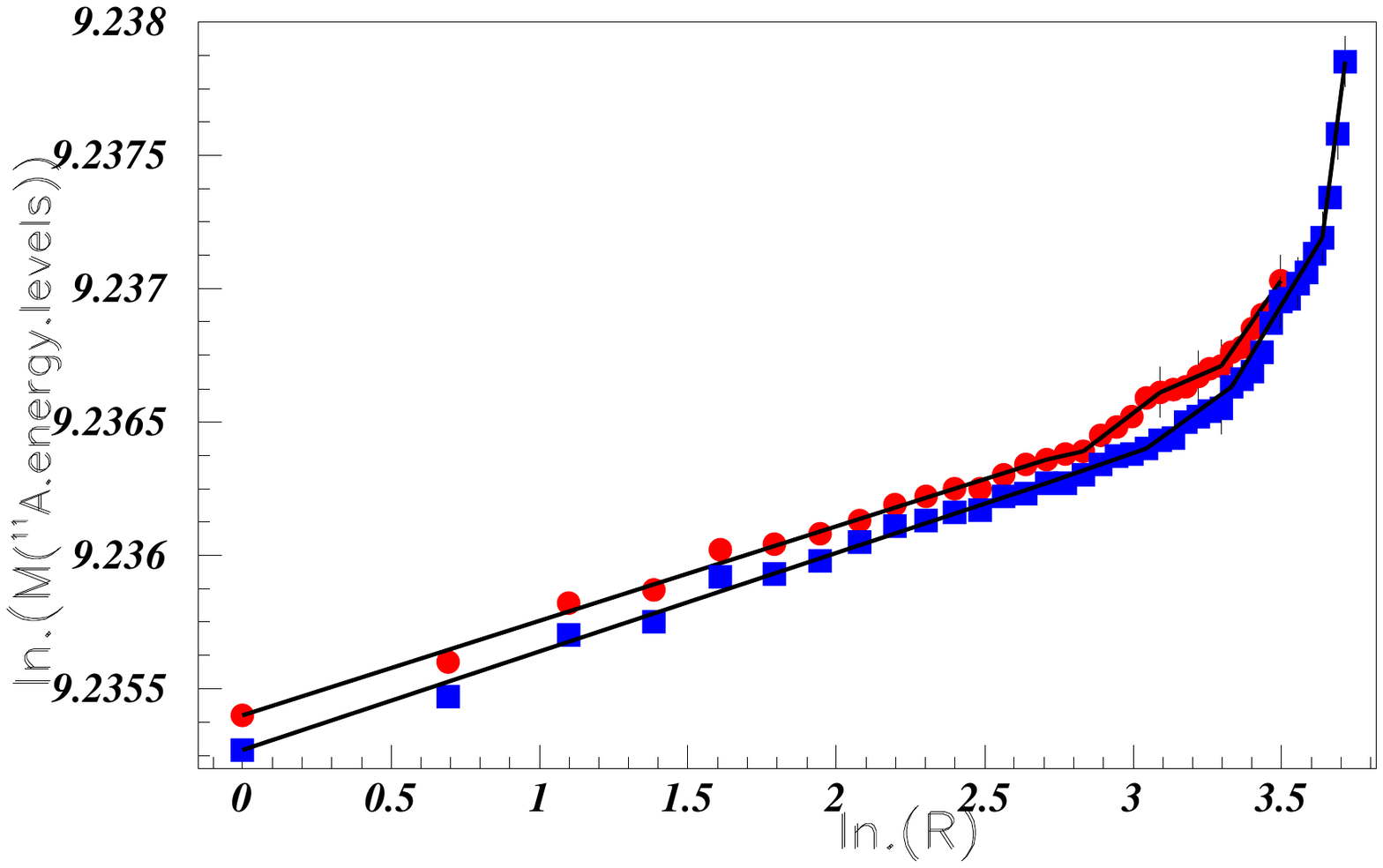}}
\end{figure}
\begin{figure}[h]
\begin{center}
\caption{Ratios of $m_{n+1}/m_{n}$ masses of $^{11}$C energy levels: full circles (red on line), and  $^{11}$B energy levels: full squares (blue on line).} 
\hspace*{-3.mm}
\scalebox{1}[1.5]{
\includegraphics[bb=15 230 520 544,clip,width=0.45\textwidth]{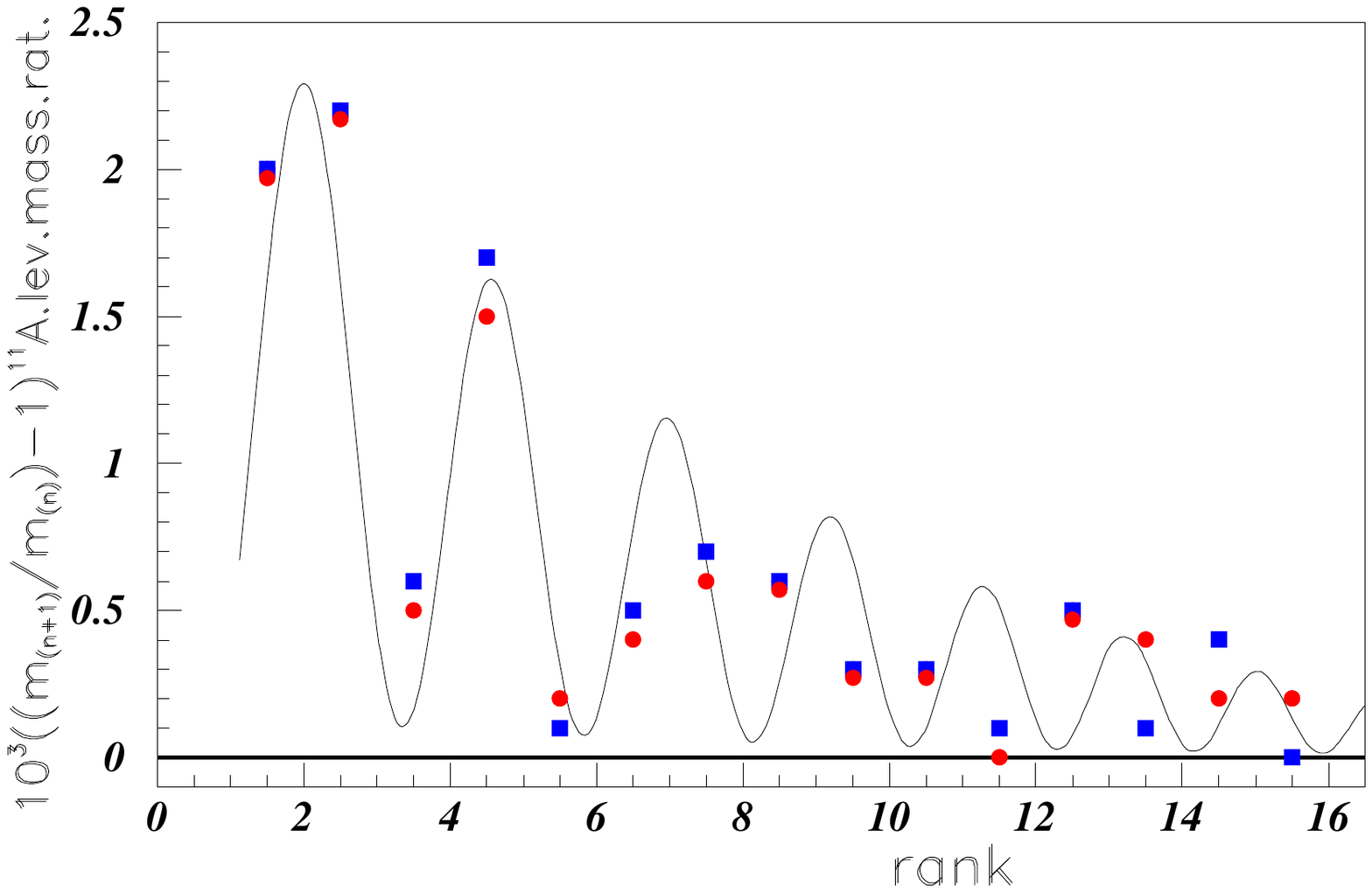}}
\end{center}
\end{figure}
\begin{figure}[h]
\caption{$^{11}$C/$^{11}$B excited energy level ratios.} 
\hspace*{-3.mm}
\scalebox{1}[1.5]{
\includegraphics[bb=6 230 530 550,clip,width=0.45\textwidth]{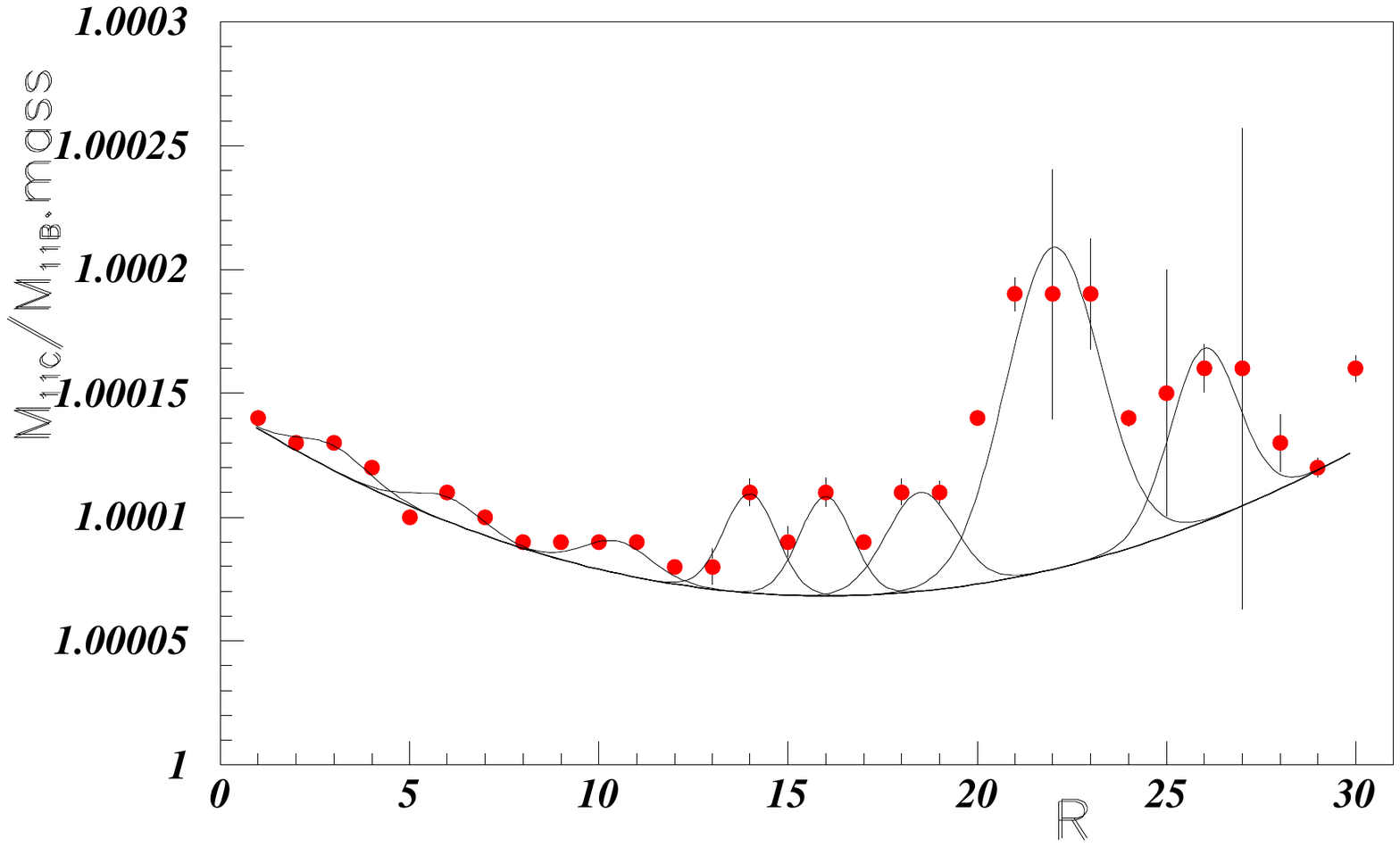}}
\end{figure}

We observe in figure~17, nice and parallel alignements of log-log distributions for the first 16 levels of $^{11}$C (full circles, red on line), and of $^{11}$B (full squares, blue on line). Figure~ 18 shows the corresponding $m_{n+1}/m_{n}$ mass ratio distributions of $^{11}$C energy levels: full circles (red on line), and  $^{11}$B energy levels: full squares (blue on line). The same set of parameters, allows, using equation (3), to well reproduce these ratios for both nuclei at least up to rank 11. 

Figure~19 shows the $^{11}$C/$^{11}$B mass ratio, fitted by arbitrary gaussians in view to emphasize the oscillating behaviour of this ratio. Equation (3) does not allow to fit these data.
\begin{figure}[ht]
\caption{Log-log plot of energy level masses of $^{12}$C.} 
\hspace*{-3.mm}
\scalebox{1}[1.5]{
\includegraphics[bb=6 230 530 550,clip,width=0.45\textwidth]{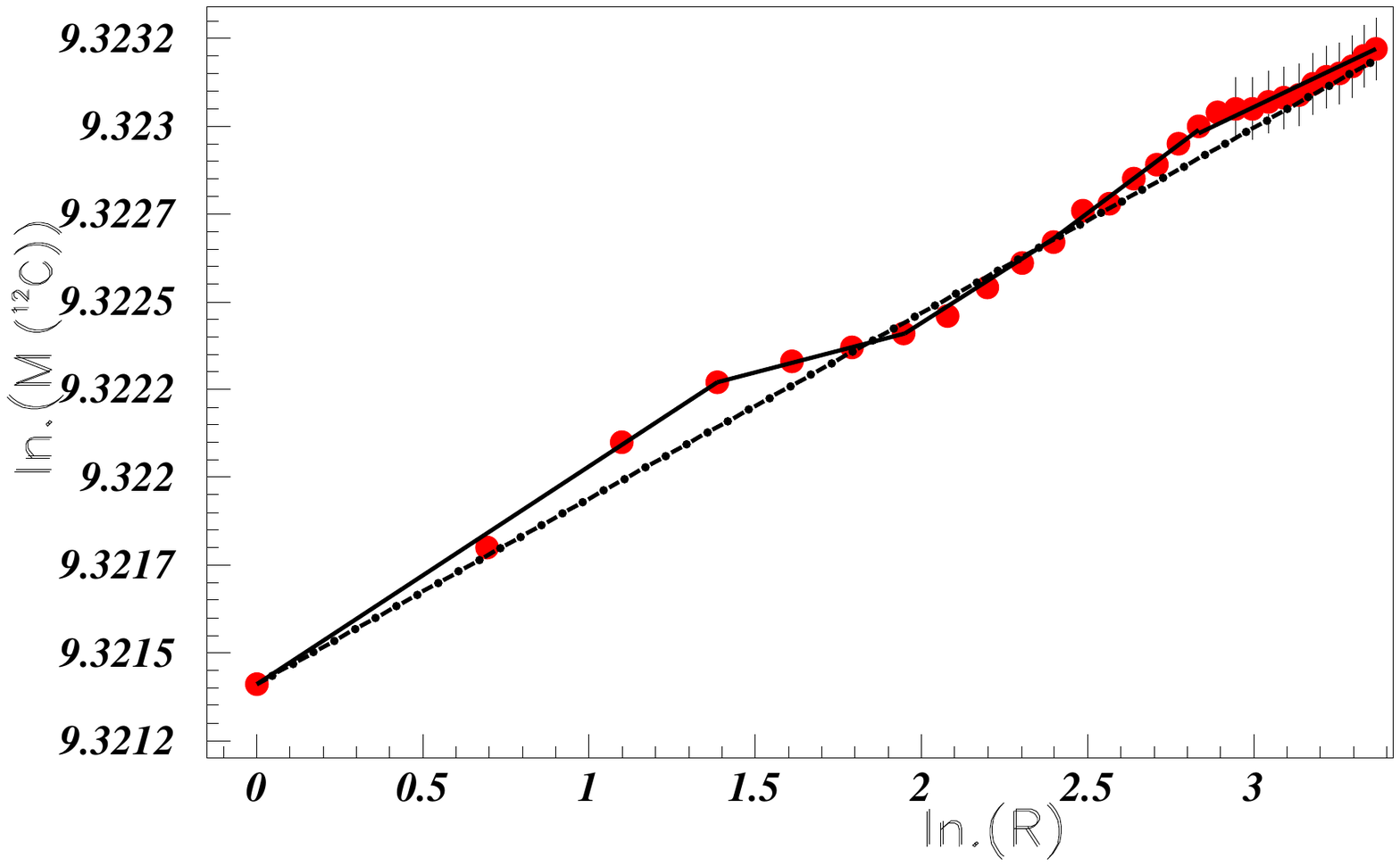}}
\end{figure}

Figure~20 shows the log-log plot of the $^{12}$C excited energy level masses \cite{as111213}.
We observe a more or less nice alignement, with two slight shifts at the beginning (low "r"), and in the vicinicy of the 17$^{th}$ level (mass = 17.76~MeV).
\begin{figure}[h]
\caption{ Ratios of $m_{n+1}/m_{n}$ masses of $^{12}$C energy levels.} 
\hspace*{-3.mm}
\scalebox{1}[1.5]{
\includegraphics[bb=12 240 520 550,clip,width=0.45\textwidth]{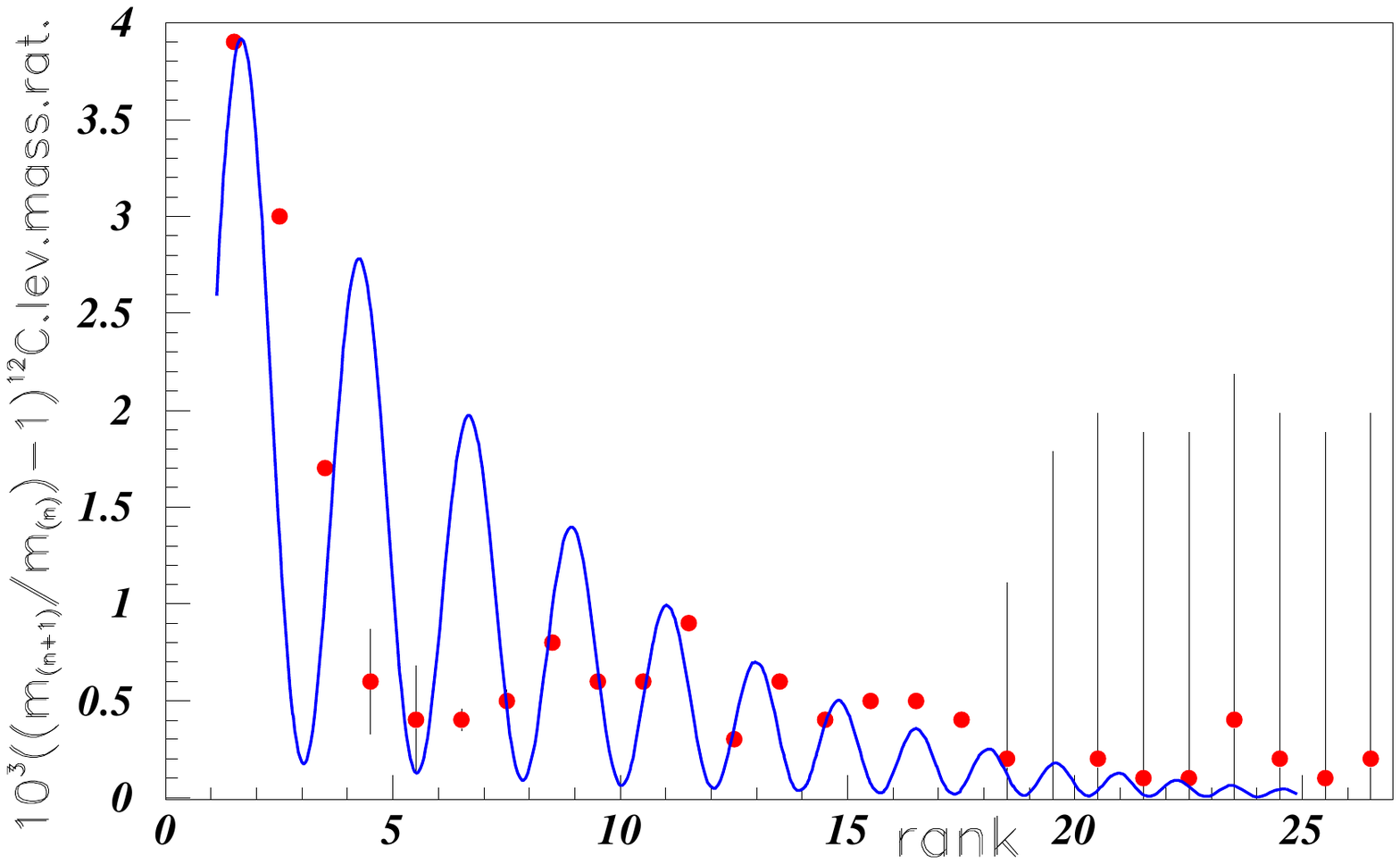}}
\end{figure}
Figure~21 shows the $m_{n+1}/m_{n}$ mass ratios between adjacent  $^{12}$C energy levels.
Since the relative mass  difference between adjacent levels is very small, a few MeV versus more than 11 GeV, The quantity 1000*((m$_{n+1}$/m$_{n}$)-1) is plotted instead of m$_{n+1}$/m$_{n}$. When the mass precision is not given in the tables, an unprecision of 1~MeV is arbitrarily introduced for these uncertainties. The large error bars starting at 
R = 17.5 manifest these large error arbitrarily introduced. The distribution fits all the fifteen first excited level masses, but the 2$^{nd}$, 3$^{rd}$, and 4$^{th}$ calculated maximas have no experimental counterpart. The ratio of rank 6.5 is far from the maximum of the calculated distribution at the same rank.

Figure~22 shows the log-log plot for A = 13 excited level masses \cite{as111213}. Full circles (red on line) show the plot for $^{13}$C nuclei, full squares (blue on line) show the plot  for $^{13}$N nuclei,  full stars (green on line) show the plot  for $^{13}$B nuclei,  and  full  triangles (purple on line) show the plot  for $^{13}$O nuclei. The alignement of these lines, is not as good as it was for the $^ {11}$C and $^ {11}$B nuclei. The  corresponding $m_{n+1}/m_{n}$ mass ratios, shown in figure~23,  are however well described by the DSI formula at least up to rank~16.
\begin{figure}[ht]
\caption{Log-log distribution of A = 13 excited level masses. Full circles (red on line) show the plot for $^{13}$C nuclei, full squares (blue on line) show the plot  for $^{13}$N nuclei,  full stars (green on line) show the plot  for $^{13}$B nuclei,  and  full  triangles (purple on line) show the plot  for $^{13}$O nuclei.} 
\hspace*{-3.mm}
\scalebox{1}[1.5]{
\includegraphics[bb=5 230 525 550,clip,width=0.45\textwidth]{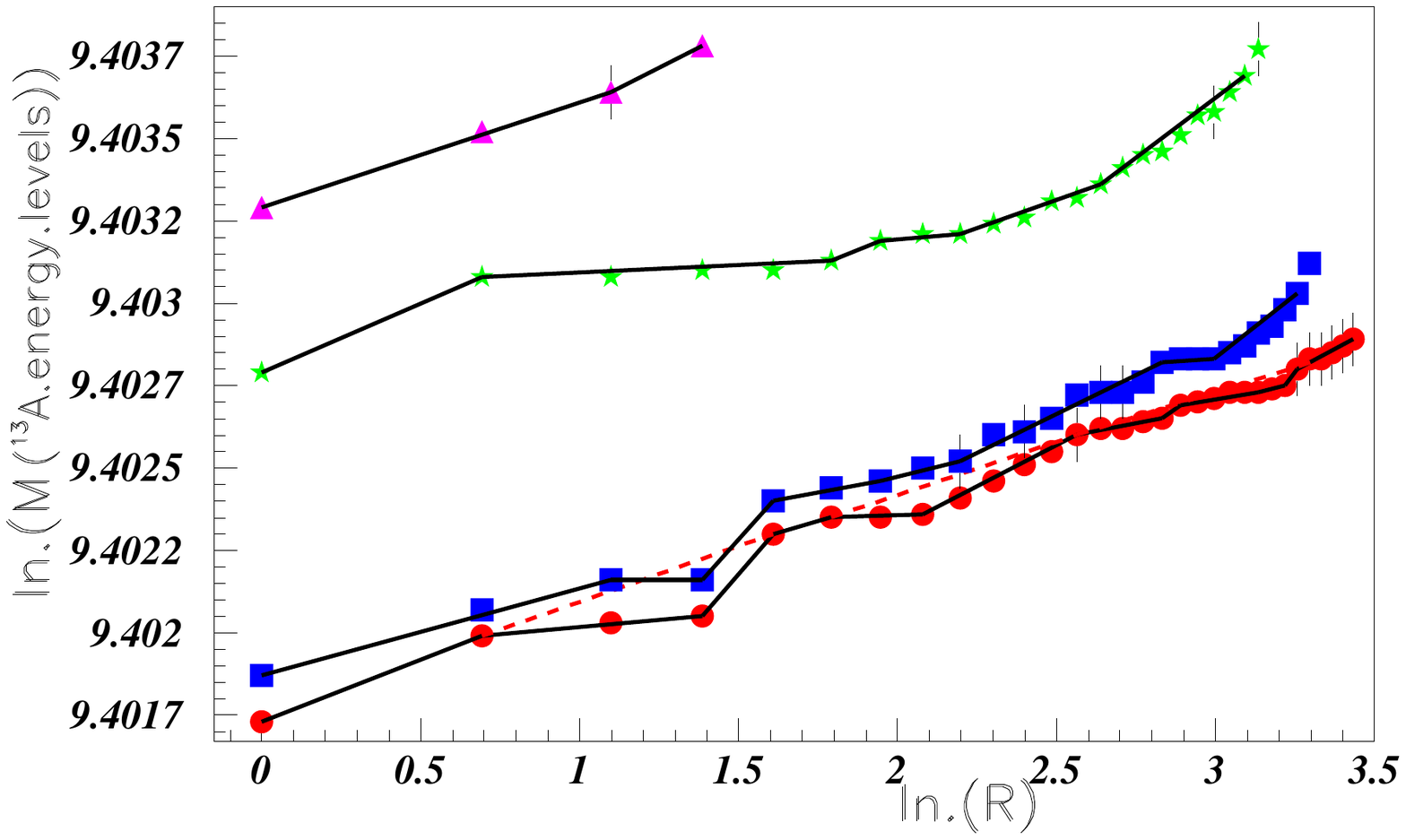}}
\end{figure}
\begin{figure}[h]
\caption{ Ratios of $m_{n+1}/m_{n}$ masses of excited levels of A = 13 nuclei. Full circles (red on line) show the distribution for $^{13}$C nuclei, full squares (blue on line) show the distribution  for $^{13}$N nuclei,  full stars (green on line) show the distribution  for $^{13}$B nuclei,  and  full  triangles (purple on line) show the distribution  for $^{13}$O nuclei.} 
\hspace*{-3.mm}
\scalebox{1}[1.5]{
\includegraphics[bb=5 230 525 550,clip,width=0.45\textwidth]{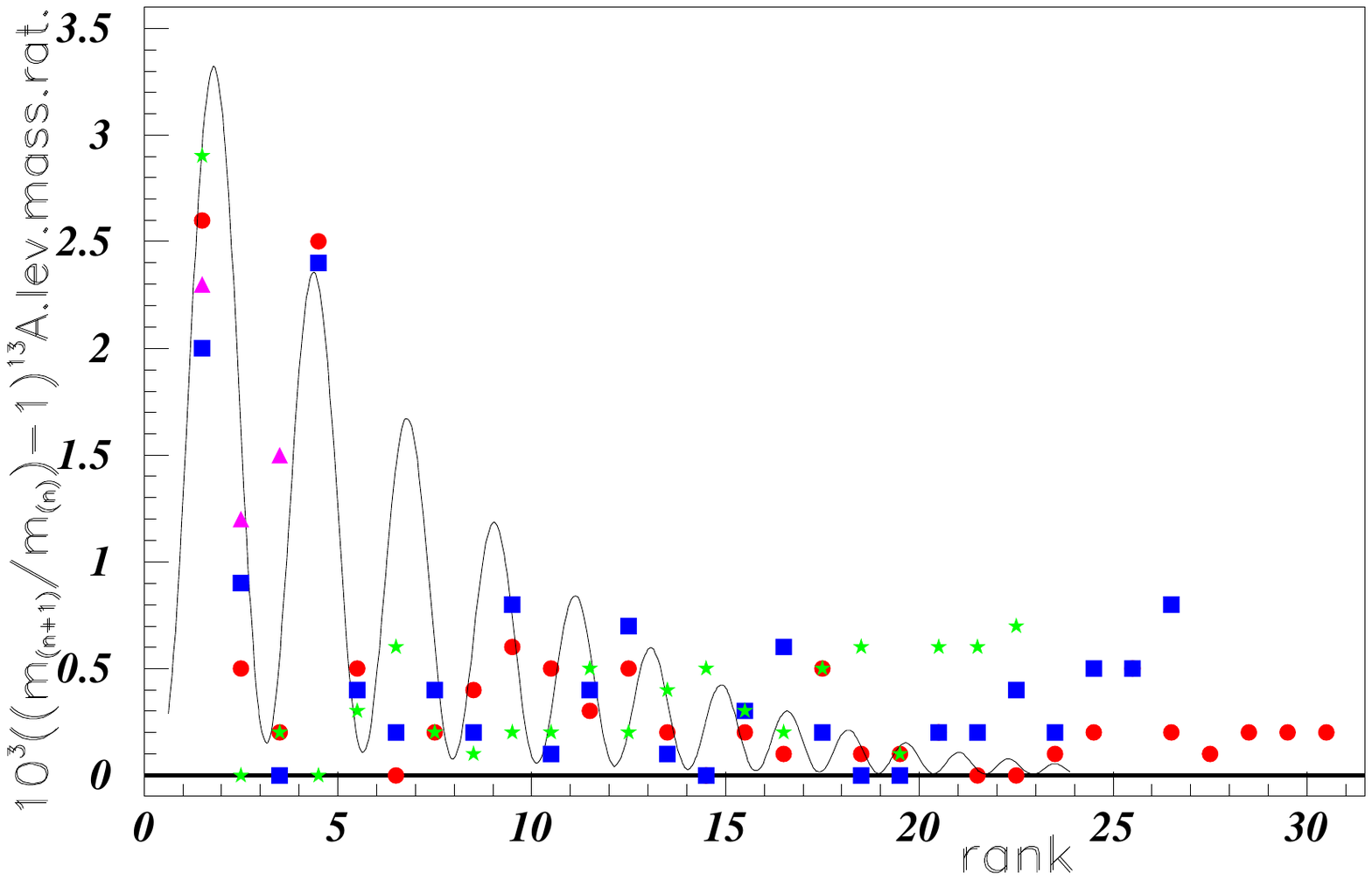}}
\end{figure}
\begin{figure}[h]
\caption{Log-log plot of energy level of $^{14}$C.} 
\hspace*{-3.mm}
\scalebox{1}[1.5]{
\includegraphics[bb=7 236 517 545,clip,width=0.45\textwidth]{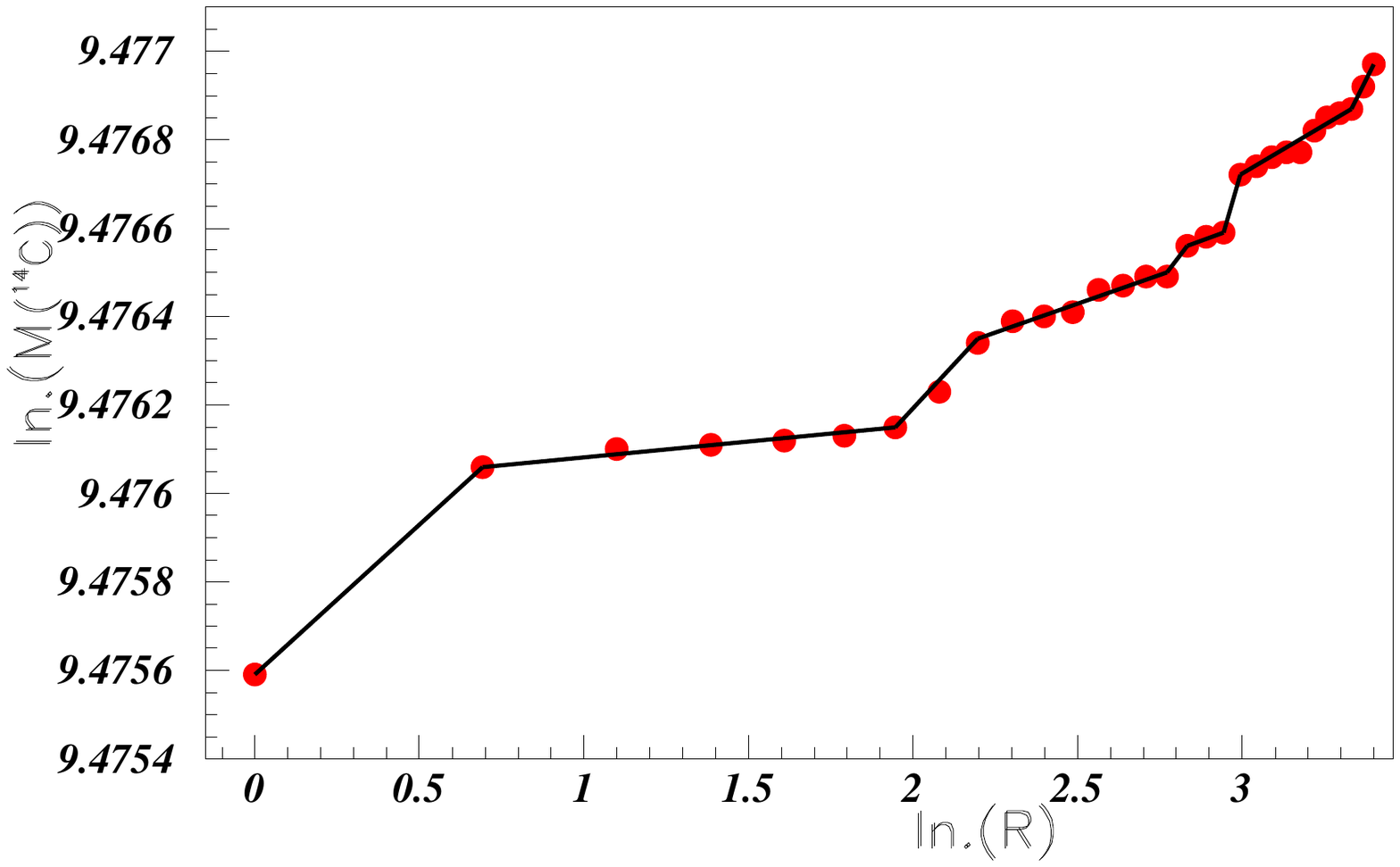}}
\end{figure}
\begin{figure}[h]
\caption{ Ratios of $m_{n+1}/m_{n}$ masses of $^{14}$C.} 
\hspace*{-3.mm}
\scalebox{1}[1.5]{
\includegraphics[bb=14 235 520 550,clip,width=0.45\textwidth]{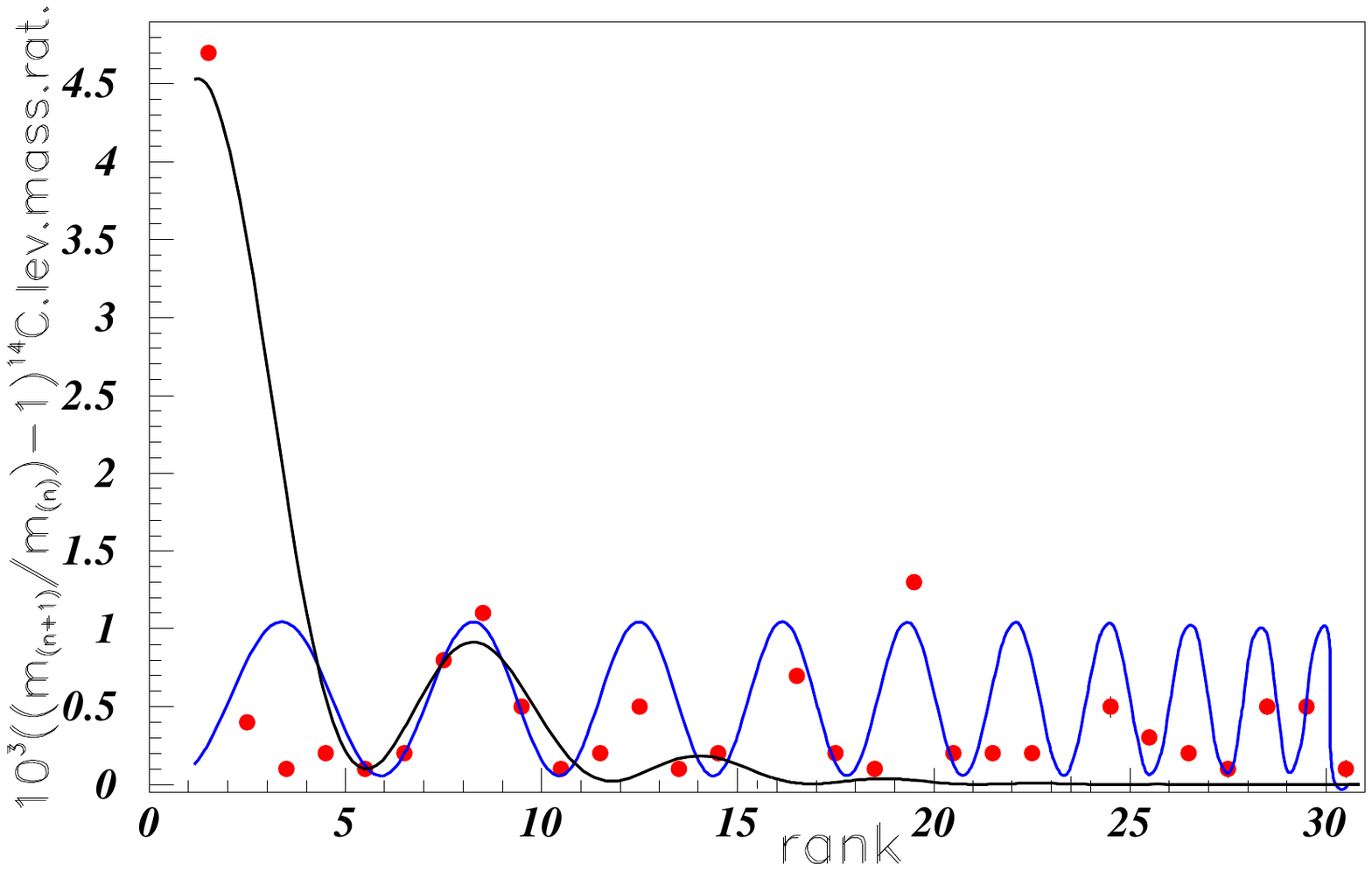}}
\end{figure}
\begin{figure}[h]
\caption{Log-log plot of energy level masses of $^{16}$O nucleus (see text).} 
\hspace*{-3.mm}
\scalebox{1}[1.5]{
\includegraphics[bb=6 135 520 550,clip,width=0.45\textwidth]{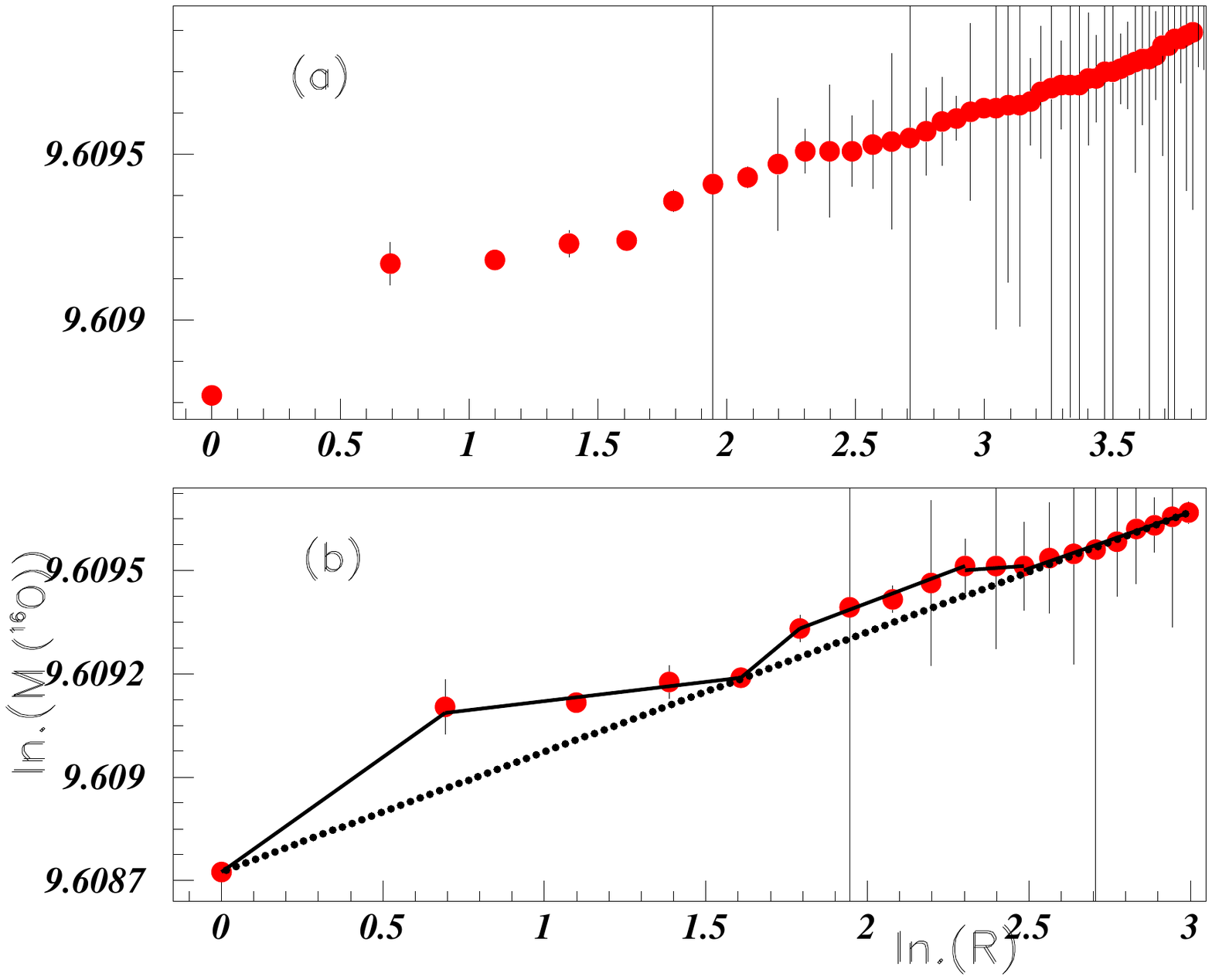}}
\end{figure}
Figure~23 shows the $m_{n+1}/m_{n}$ mass ratios of excited levels for 
A~=~13 nuclei. The curve corresponds to a fit obtained using  equation (3).
As before, we observe an unique and rather good fit for all four A = 13 nuclei, up to R~=~16, which spoils for larger R values. 
\subsection{Application to $^ {14}$C energy levels}
Figure~24 shows the log-log plot of $^{14}$C energy level masses \cite{as14}; several straight lines are observed.

Figure~25 shows the $m_{n+1}/m_{n}$ mass ratios of $^{14}$C levels.
We observe here a case where the equation (3) does not allow to fit the experimental data. Two fits are drawn, both describe well the "peak" around rank 8. The first fit (black on line), catches also the first large point, but forgets both the data after rank 11 and the data for rank 2 and 3.  The second fit (blue on line) forgets completely the first four data points. None are satisfactory.
\subsection{Application to $^ {16}$O energy levels}
Figure~26 shows the log-log plot for $^ {16}$O nuclei level masses \cite{as16}. A large number of nuclear level masses are given in the table, but their definition spoils for energy levels with an excitation energy larger than 
$\Delta$M=13~MeV (see insert (a)). Therefore a reduced number of log-log distributions of level masses is represented in figure 26(b) .
A nice alignement is disturbed between 6 and 7~MeV excitation energy, and in the region close to 9 - 10~MeV excitation energy.
\begin{figure}[ht]
\caption{ Ratios of $m_{n+1}/m_{n}$ excited level masses of $^{16}$O.} 
\hspace*{-3.mm}
\scalebox{1}[1.5]{
\includegraphics[bb=14 235 520 550,clip,width=0.45\textwidth]{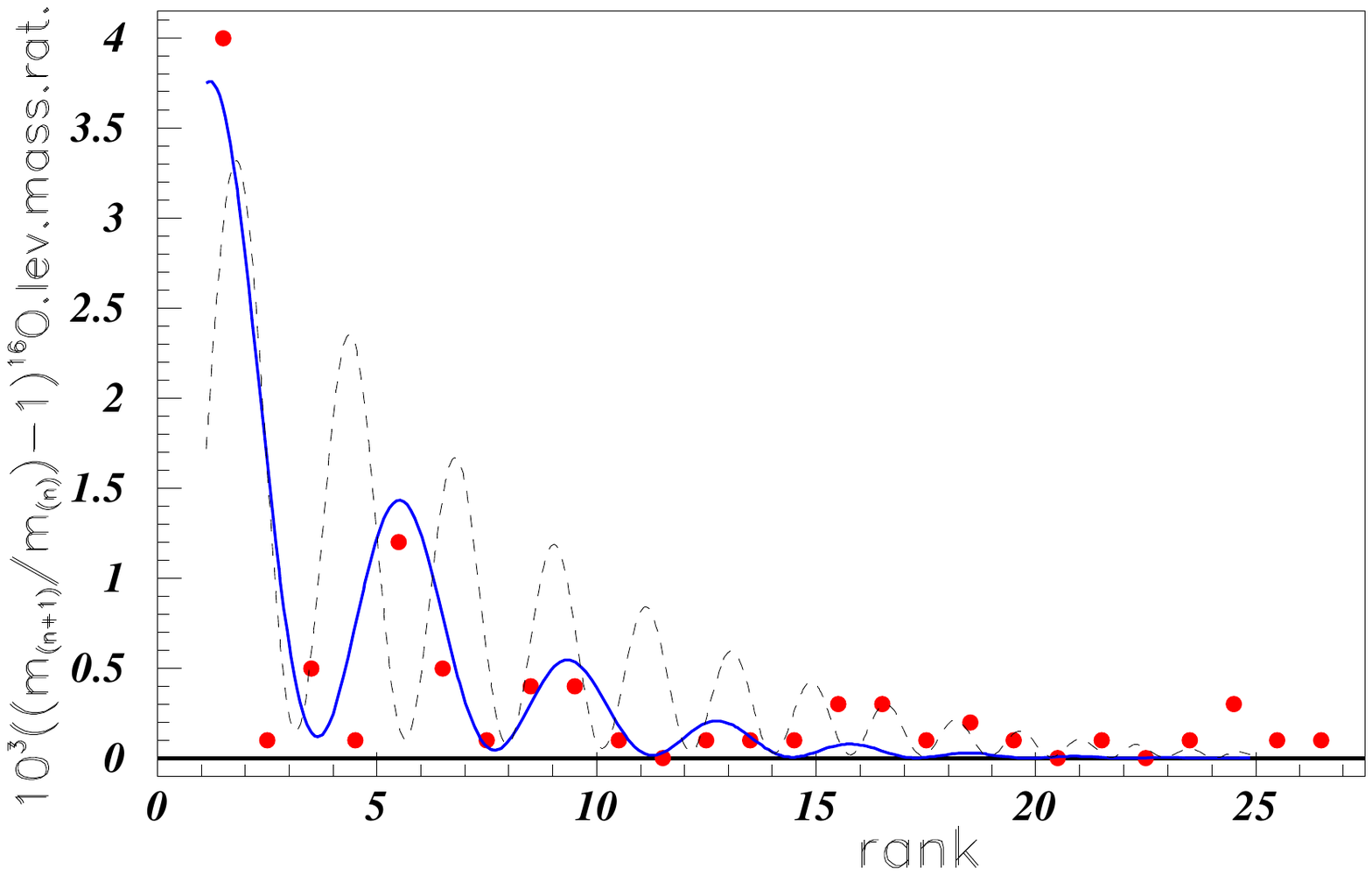}}
\end{figure}
Figure~27 shows the $m_{n+1}/m_{n}$ excited level mass ratios of $^{16}$O levels. The data points are reasonably fitted by  equation (3) up to R=14.5, but with different parameters as those used for A = 11, 12, and 13  shown in previous figures. In the figure~27, the dashed curve shows the fit obtained when the parameters fitting the C = 13 nuclei, are used. \\
\subsection{Application to $^{23}$Na nucleus}
\begin{figure}[ht]
\caption{Log-log plot of $^{23}$Na  energy level masses.} 
\hspace*{-3.mm}
\scalebox{1}[1.5]{
\includegraphics[bb=6 234 516 543,clip,width=0.45\textwidth]{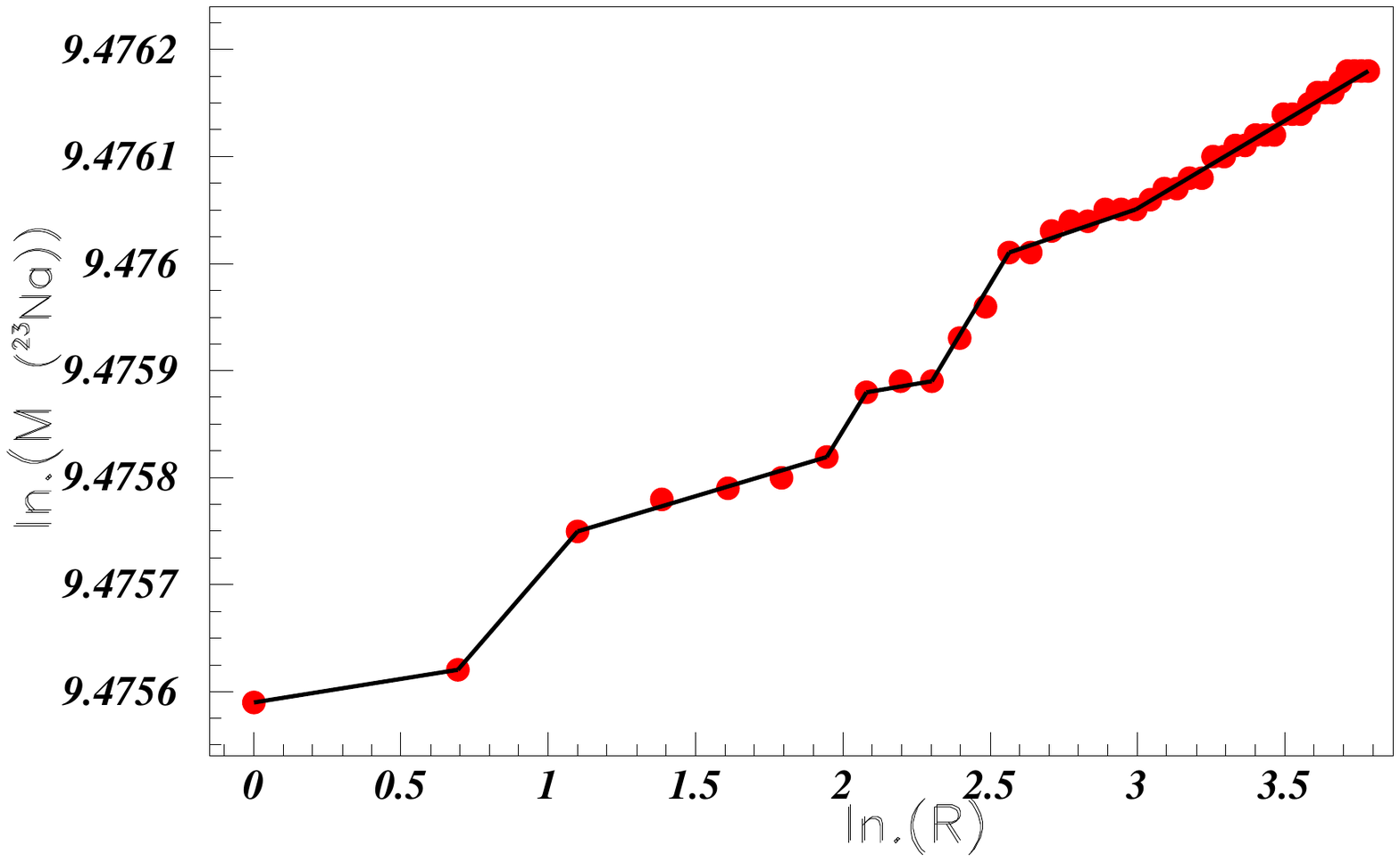}}
\end{figure}
\begin{figure}[h]
\caption{ Ratios of $m_{n+1}/m_{n}$ masses of $^{23}$Na energy levels.} 
\hspace*{-3.mm}
\scalebox{1}[1.5]{
\includegraphics[bb=14 235 520 550,clip,width=0.45\textwidth]{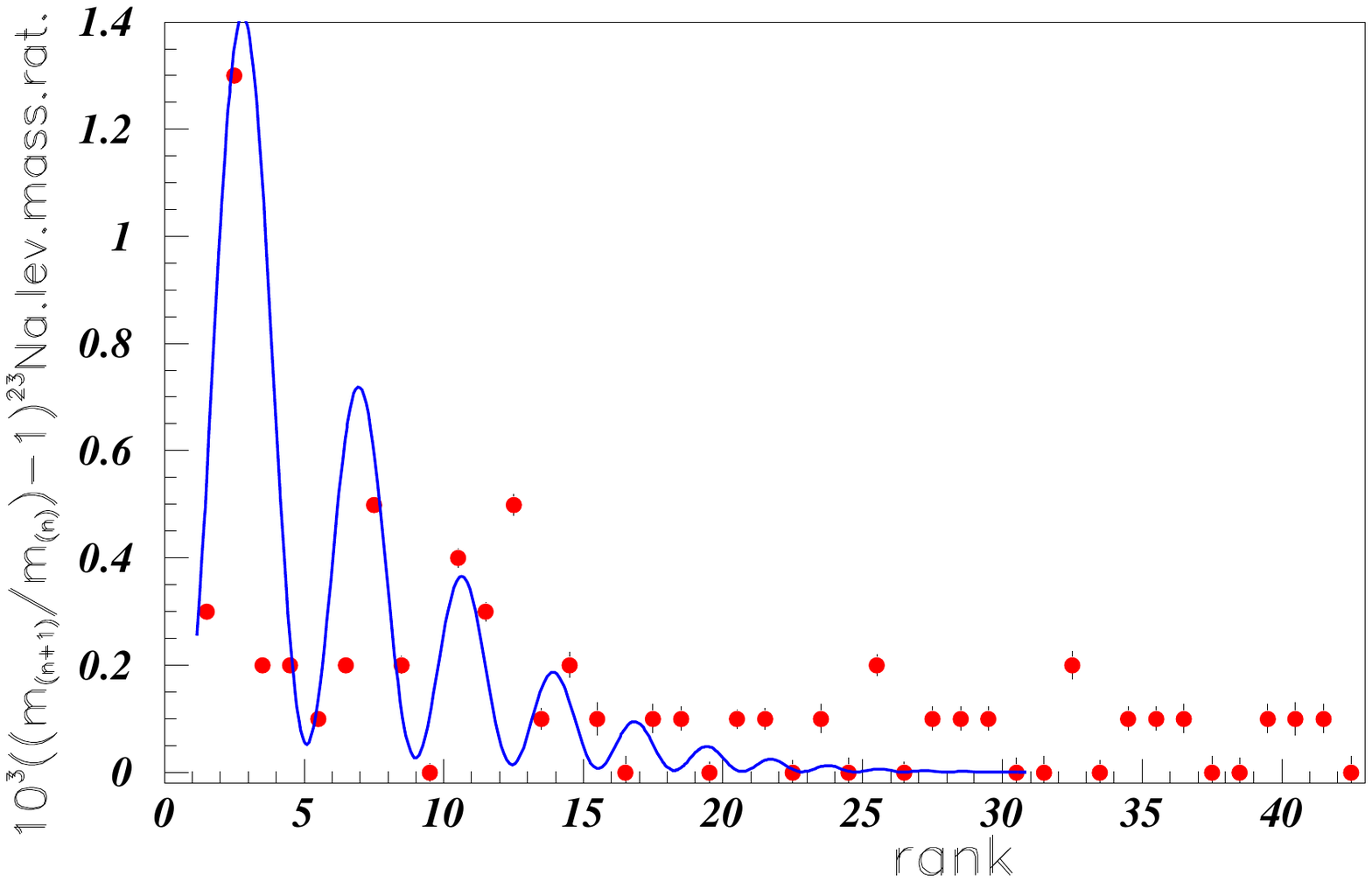}}
\end{figure}
Figure~28 shows the log-log distribution of the energy levels of the $^{23}$Na nucleus \cite{23na}. It exhibits several, too short straight lines. The  corresponding $m_{n+1}/m_{n}$ excited level mass ratio distributions, shown in figure~29, display an oscillatory pattern, rather well described by equation (3) up to rank 11.
\subsection{Application to $^{46}$Ti nucleus}
\begin{figure}[h]
\caption{Log-log plot of energy levels of $^{46}$Ti.} 
\hspace*{-3.mm}
\scalebox{1}[1.5]{
\includegraphics[bb=20 235 513 542,clip,width=0.45\textwidth]{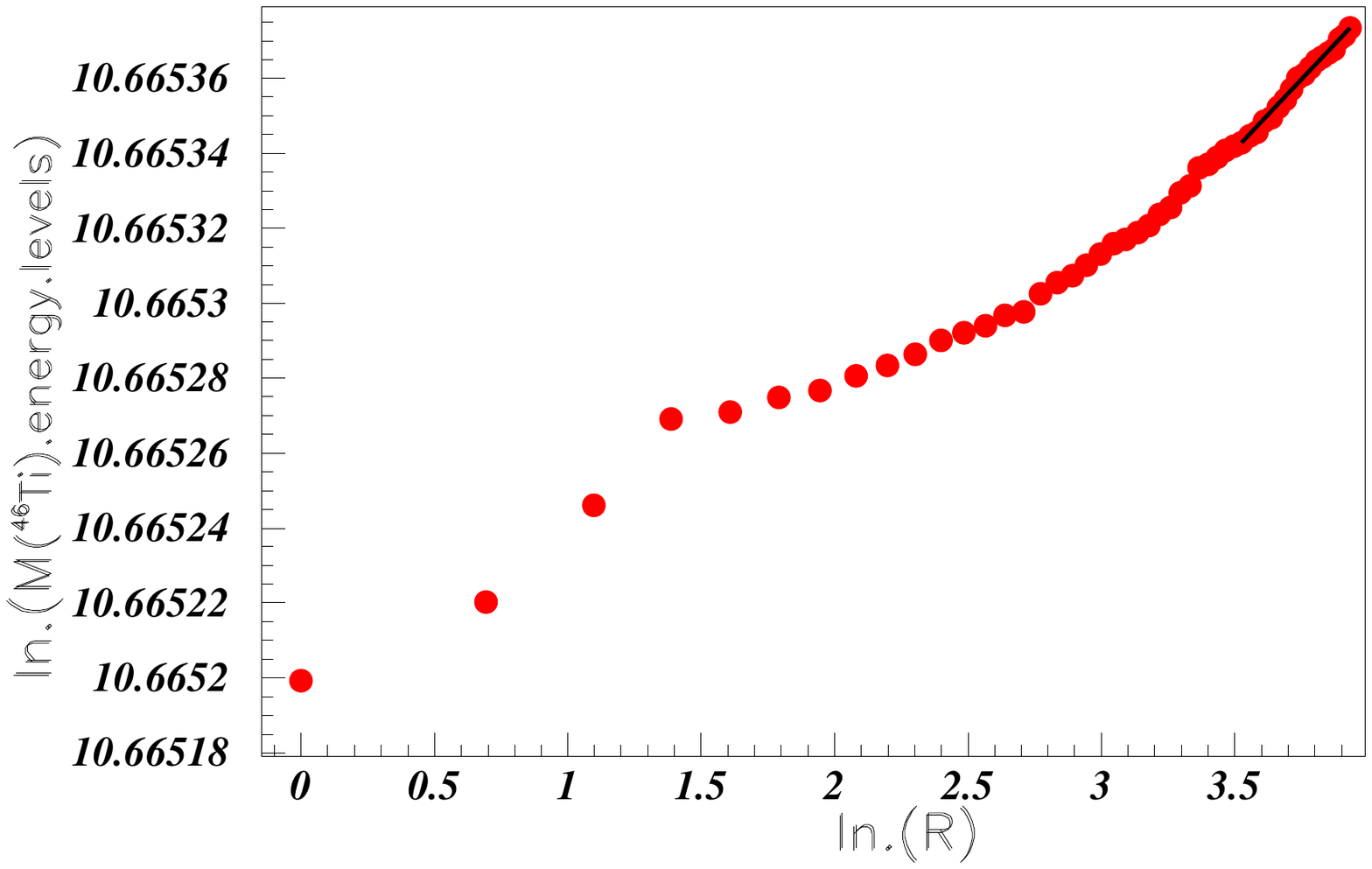}}
\end{figure}
Figure~30 shows the log-log distribution of the energy levels of the $^{46}$Ti nucleus \cite{46ti}. The linearity between the log of the masses versus the log of the rank, is not really observed, at least for the first 14 masses.  Therefore the corresponding $m_{n+1}/m_{n}$ mass ratio is not studied.
\subsection{Application to $^{62}$Ni nucleus}
\begin{figure}[h]
\caption{Log-log plot of energy level of $^{62}$Ni.} 
\hspace*{-7.mm}
\scalebox{1}[1.4]{
\includegraphics[bb=45 235 513 542,clip,width=0.45\textwidth]{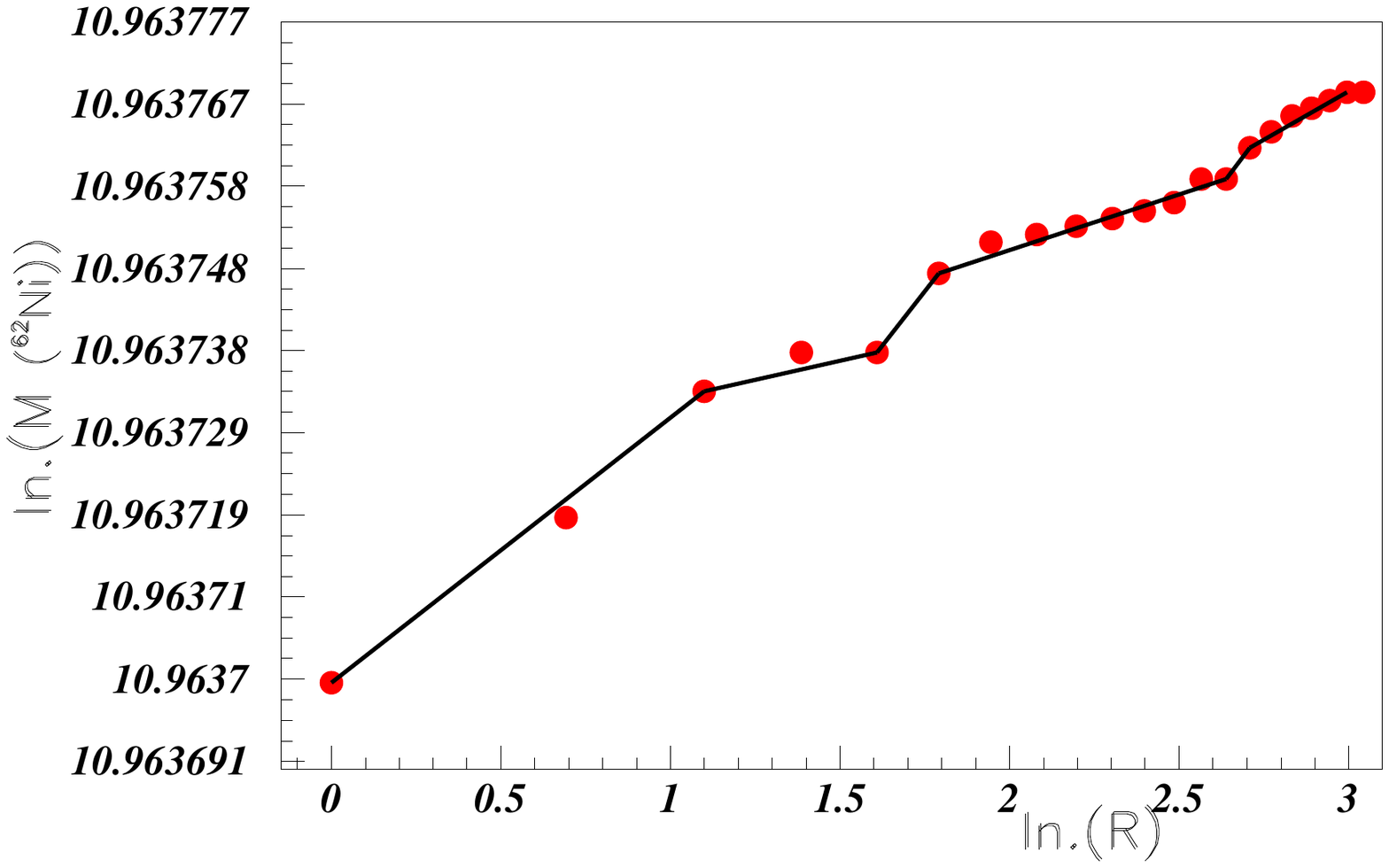}}
\end{figure}
\begin{figure}[ht]
\caption{ Ratios of $m_{n+1}/m_{n}$ masses of $^{62}$Ni.} 
\hspace*{-7.mm}
\scalebox{1}[1.4]{
\includegraphics[bb=22 233 517 542,clip,width=0.45\textwidth]{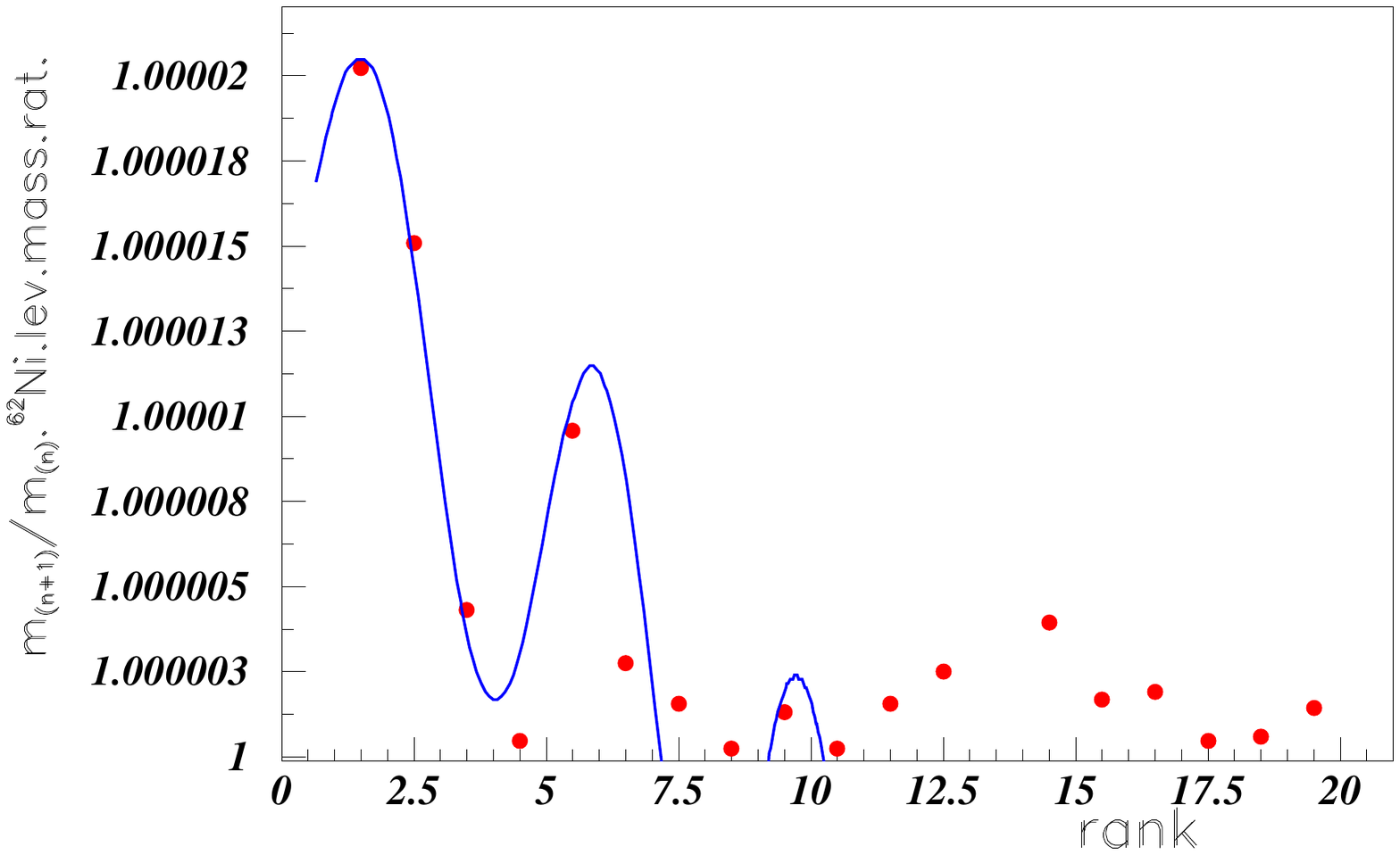}}
\end{figure}
Figure~31 shows the log-log distribution of the energy levels of the $^{62}$Ni nucleus \cite{62ni}.The distribution
 exhibits several,  straight lines, except the fourth excited level mass, M = 2.336~MeV, which is outside the linearity. 
 
The  corresponding $m_{n+1}/m_{n}$ excited level mass ratio distribution, shown in figure~32 displays an oscillatory pattern, well described by equation (3) up to rank 11.  Here also, the very close to one value of the ratio between two adjacent masses, leads us to keep only the first term of the limited development of the ln(1+x) function. 
\subsection{Application to $^{92}$Zr excited level masses}
\begin{figure}[ht]
\caption{Log-log plot of $^{92}$Zr of energy level masses.} 
\hspace*{-27.mm}
\scalebox{1}[1.45]{
\includegraphics[bb=13 236 515 542,clip,width=0.45\textwidth]{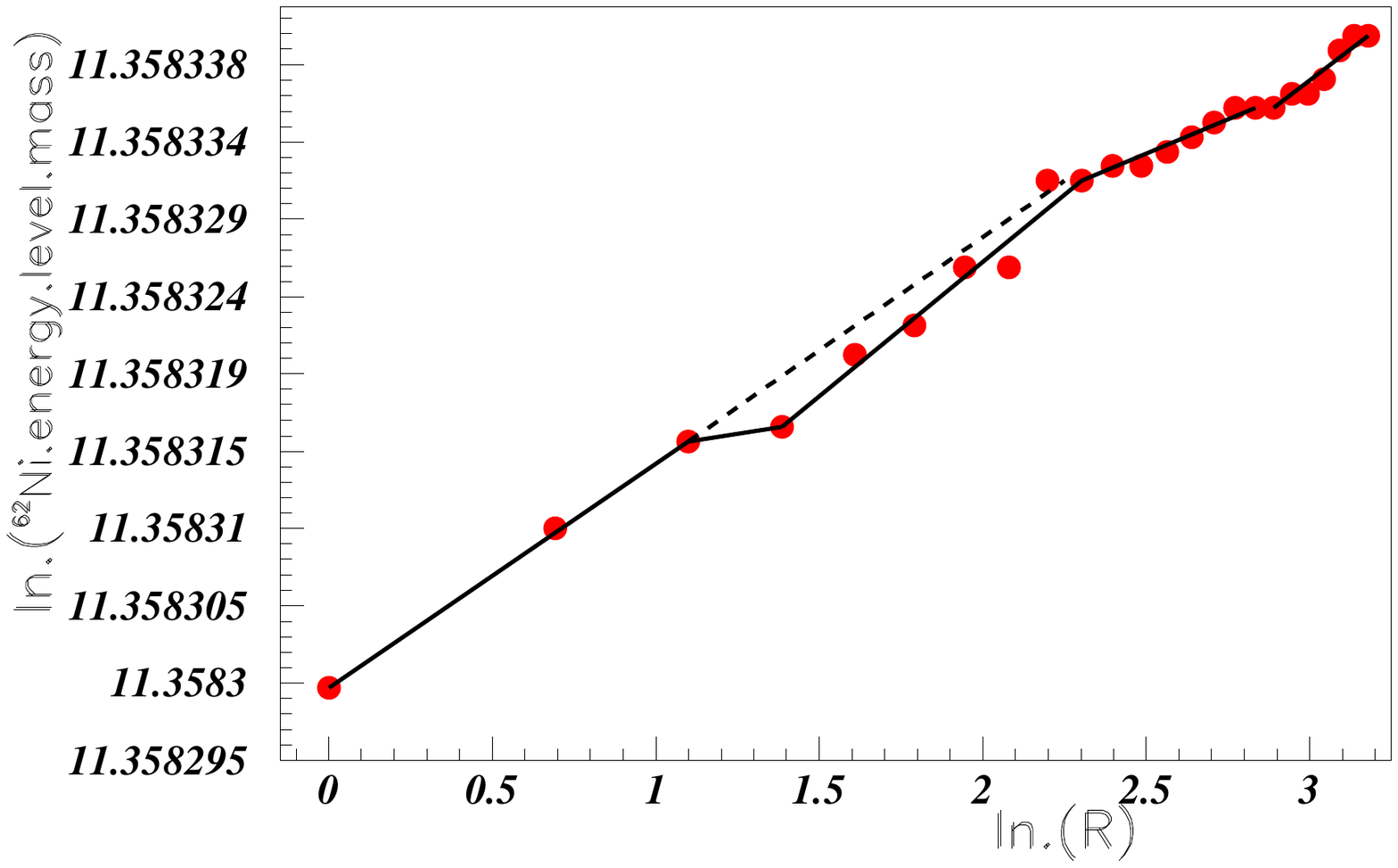}}
\end{figure}
\begin{figure}[h]
\caption{ Ratios of $m_{n+1}/m_{n}$ masses of $^{92}$Zr.} 
\hspace*{-12.mm}
\scalebox{1}[1.45]{
\includegraphics[bb=14 235 515 545,clip,width=0.45\textwidth]{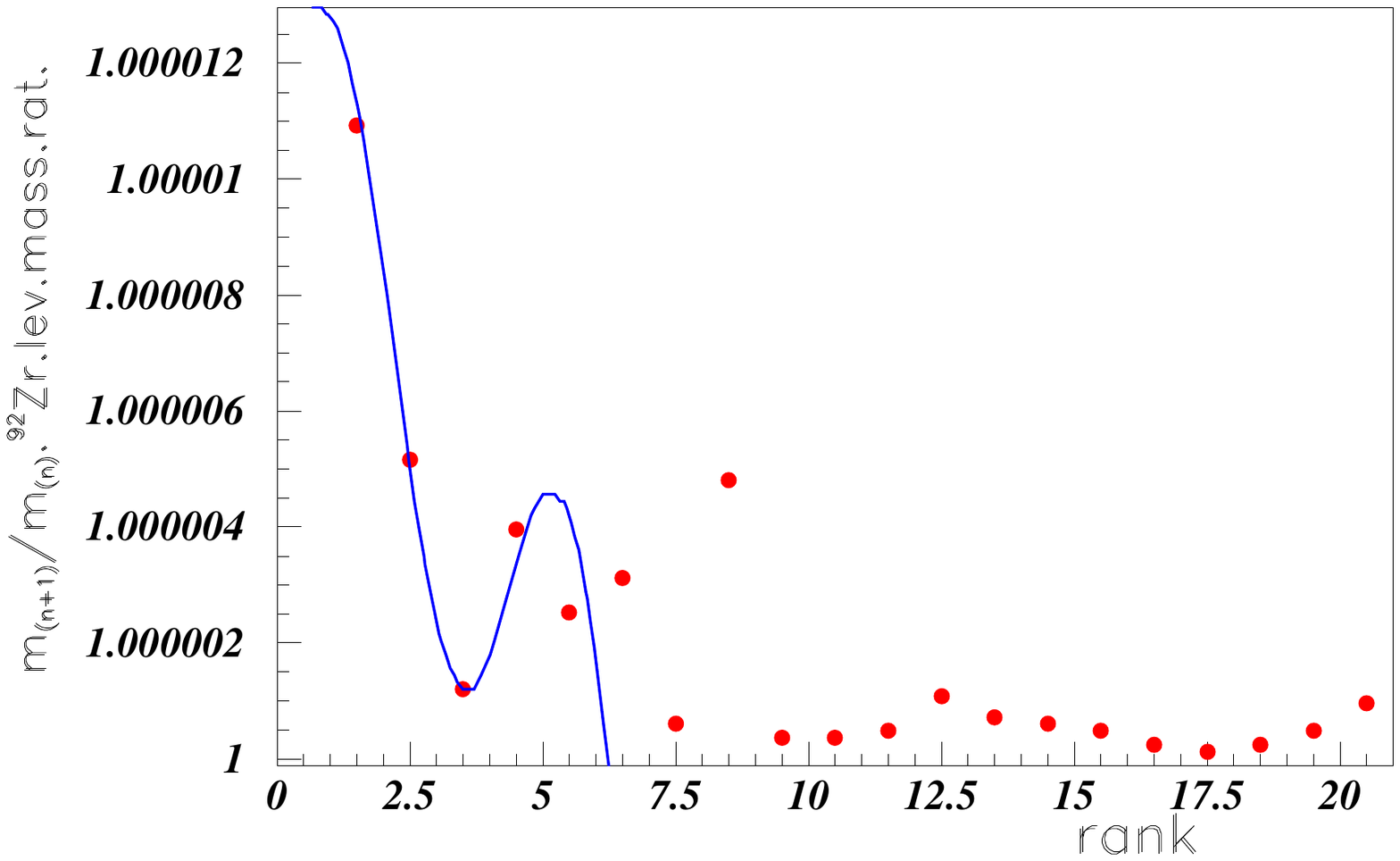}}
\end{figure}
Figure~33 shows the log-log distribution of the energy levels of the $^{92}$Zr nucleus \cite{92zr}.  The distribution exhibits some straight lines.
 
The  corresponding $m_{n+1}/m_{n}$ mass ratio distribution, shown in figure~34, displays an oscillatory pattern, well described by equation (3) but only up to rank 6, in a much smaller range than before. Here also, the very close to one value of the ratio between two adjacent masses, leads to keep only the first term of the limitated development of the ln(1+x) function.
 \subsection{Application to $^{134}$Ba the energy level masses}
\begin{figure}[ht]
\caption{Log of the energy level masses of the $^{134}$Ba nucleus versus the log of the rank "R"}
\hspace*{-3.mm}
\scalebox{1}[1.45]{
\includegraphics[bb=10 240 520 544,clip,width=0.45\textwidth]{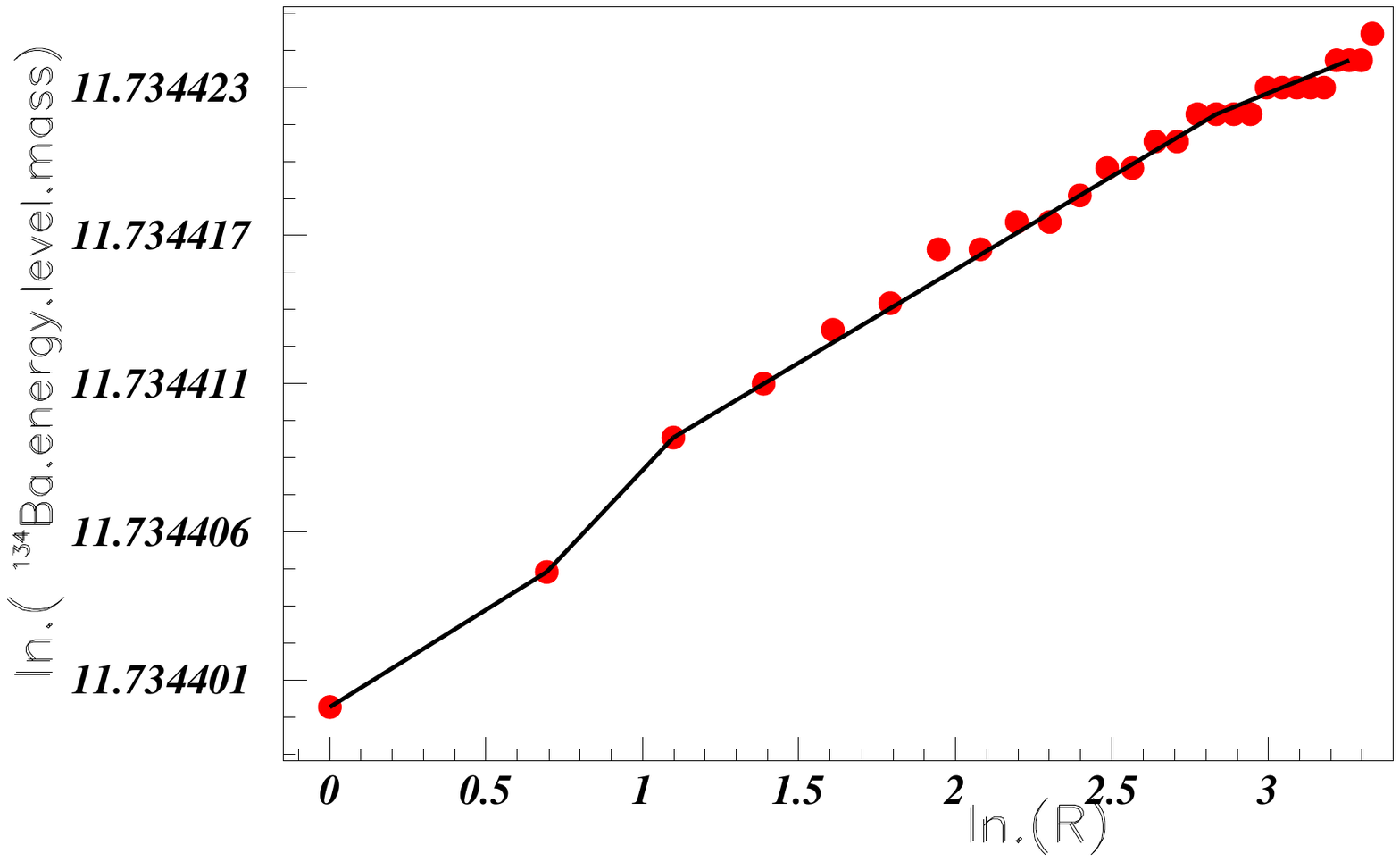}}
\end{figure}
\begin{figure}[ht]
\caption{Ratios of $m_{n+1}/m_{n}$ masses for $^{134}$Ba nucleus}
\hspace*{-3.mm}
\scalebox{1}[1.45]{
\includegraphics[bb=7 240 515 544,clip,width=0.45\textwidth]{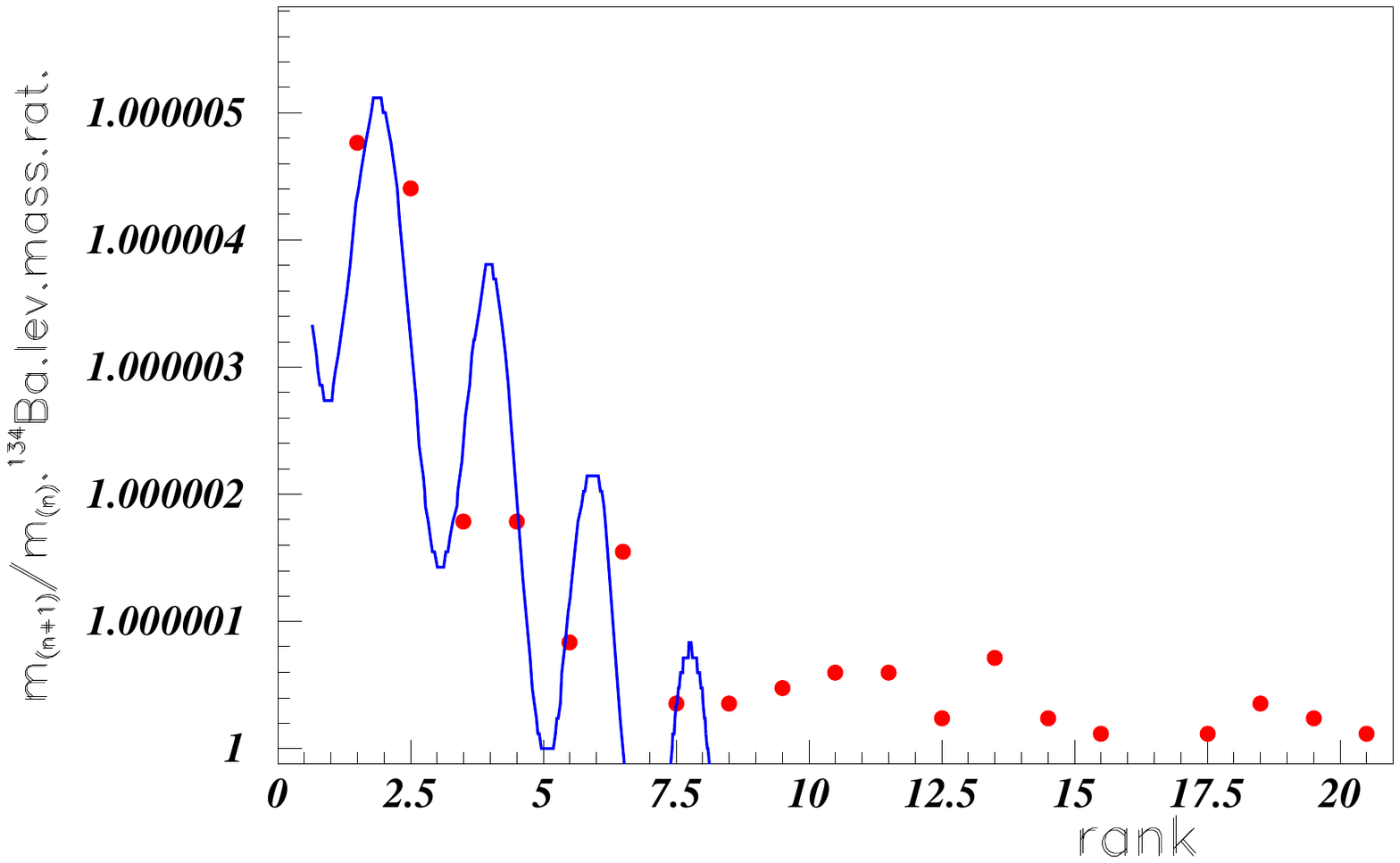}}
\end{figure}
Figure~35 shows the log-log distribution corresponding to $^{134}$Ba nucleus energy level masses \cite{134ba}. A long straight line is observed for this nuclei; a nice fit is also observed, up to rank R~=~8, for the  $^{134}$Ba $m_{n+1}/m_{n}$ mass ratio versus the rank "R", shown in figure~36.
\section{Discussion}
The values extracted from the fits analysing the mass ratios, are given in table~1. 
Although, it is not possible to give the precisions on these parameters, we observe that an unique curve, describing the data, is plotted in figures 18 and 23, although the parameters shown in table~III, differ a little. This gives an idea of the parameter uncertainty. The data differ from fit, always beyond R~=~16, sometimes before.

The parameters, obtained from section II, studying the nuclei mass variations, describe the complete distributions and are therefore stronger than those determined from excited level energies (section III), for which only the beginning of the distributions are well fitted. We observe that the range of linearity in the log-log distribution, does not always correspond to the range of nice fit, between the mass ratio data and the curve obtained using equation (3).

Considering the parameters obtained from nuclei masses, we observe:\\
 - all $\Omega$ values are close, except that from A~=~251 which is a little lower: 14 instead of 17. Therefore the $\lambda$ and                 
 Im($\alpha$) are also almost constant,\\
 - $a_{1}$ and Re($\alpha$) decrease for increasing masses A.\\ 
 - all $\mu$ are equal or close to 1.\\
 
Considering the parameters obtained from nuclei excited levels, we observe:\\
 - two distinct $\Omega$, therefore $\lambda$ and Im($\alpha)$ values, have been extracted. The first one, close to 14.2,   is again a little lower than $\Omega$~=~17, observed in the nuclei masses studies (except for $^{134}$Ba , figure~36). The second value is $\Omega$~$\approx$8.2. The transition occurs between masses A~=~13 and A~=~14,\\
  - the $a_{1}$ parameters are much larger than before, except for the two heavier nuclei studied: $^{92}$Zr and $^{134}$Ba, where it takes the same value as previously in  heavy nuclei mass fits, namely $a_{1}\le~10^{-4}$.\\
  - $\mu$ equals 1  three times, equals 1.4  seven times, equals 2  two times, and equals 2.6 one time. These number are respectively close to the square  roots of 1, 2, 4, and 7. The variation of $\mu$ is due to two opposite effects:
an increase for stable nuclei close to magic numbers, in comparison with less stable nuclei, and a decrease for larger mass nuclei, since it is the relative mass variation which is concerned. This second effect is the main one for heavy nuclei, as seen in table 1 for $^{92}$Zr and $^{134}$Ba nuclei. 
\begin{figure}
\caption{Variation of $\lambda$ values  versus A. Full triangles (mauve on line) show the values extracted from nuclei mass variations. Full stars (sky blue on line) show the values from nuclei energy level masses.}
\hspace*{-3.mm}
\scalebox{1}[1.5]{
\includegraphics[bb=27 280 525 525,clip,width=0.45\textwidth]{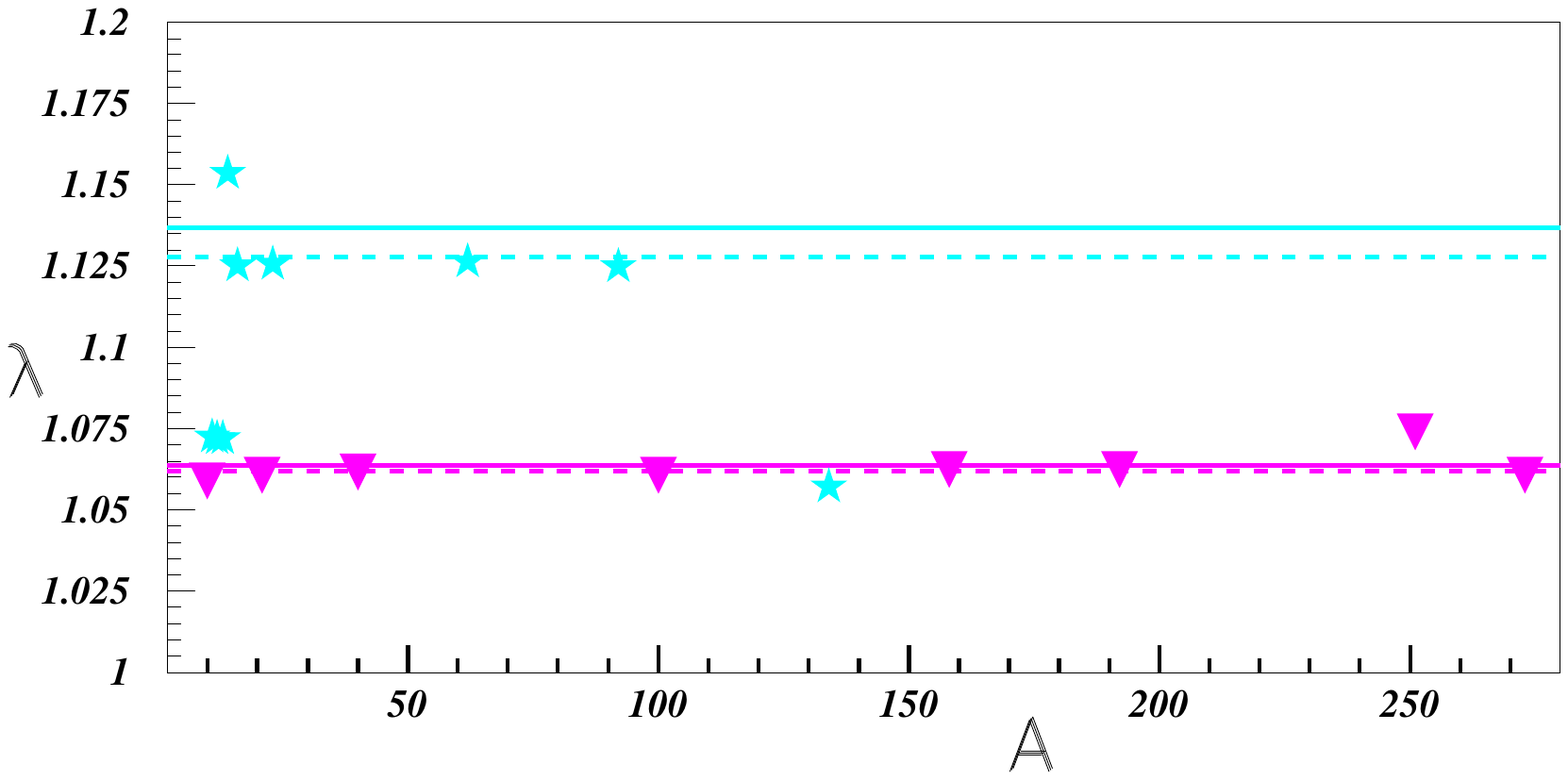}}
\end{figure}

Figure~37 shows the values of the fundamental scaling ratio $\lambda$ versus A. In figure~37, full triangles (mauve on line) show the values extracted from nuclei mass variations. These values are constant around $\lambda_{1}$~=~1.062 (dashed straight line, mauve on line) with a very small variation around this mean value. Since several fits give the same values for $\lambda$, the figure shows less points than obtained by fits.

Full stars (sky blue on line) show the values from the nuclei excited energy level masses. Four of them, extracted from light and heavy nuclei, have the same $\lambda$ value, namely $\lambda_{2}$~$\approx$~1.128, shown in figure~37 by a dashed straight line (sky blue on line). Two other $\lambda$ values are close to 1.062. Such property was anticipated in \cite{sornette} where it was noticed that  "DSI obeys scale invariance for specific choices of $\lambda$, which form an infinite set of values that can be written as $\lambda_{n}$ = $\lambda^{n}$. Indeed here  $\lambda_{2}$~=~$\lambda_{1}^{2}$.

Notice that the low $\lambda$ value is close to the $\lambda$ observed previously \cite{arxiv1104} for quarks: $\lambda$~=~1.074 and leptons: $\lambda$~=~1.061. Just as closed $\lambda$ values were extracted from the fits which analyzed in the same way the  fractal properties of mesons and baryons \cite{frachadron}. Then regrouping all $\lambda$ values obtained through fits performed on quarks, leptons, hadrons and nuclei, we get the two continuous lines drawn in figure~37, where we have shared the $\lambda$ into two groups. The first group leads to the first mean value $\bar{\lambda_{1}}$~=~1.064, using all $\lambda$~$\le$~1.1; the second group use $\lambda$~$\ge$~1.1 giving 
${\bar \lambda_{2}}$~=~1.137. We notice that we still have 
${\bar \lambda_{2}}$~=~${\bar \lambda_{1}^{2}}$.

\begin{figure}[ht]
\caption{Plot of Im($\alpha$) versus Re($\alpha$). Full circles (red on line) show the $\alpha$ values which fit the elementary particle quark and lepton mass distributions; full up side triangles (green on line) show the same for meson data; full squares (blue on line) show the same for baryon data; full down side triangles (purple on line) show the same for nuclei mass variation data; and full stars (blue sky on line) show the same for nuclei energy level mass data.}
\hspace*{-3.mm}
\scalebox{1}[1.5]{
\includegraphics[bb=4 286 520 520,clip,width=0.45\textwidth]{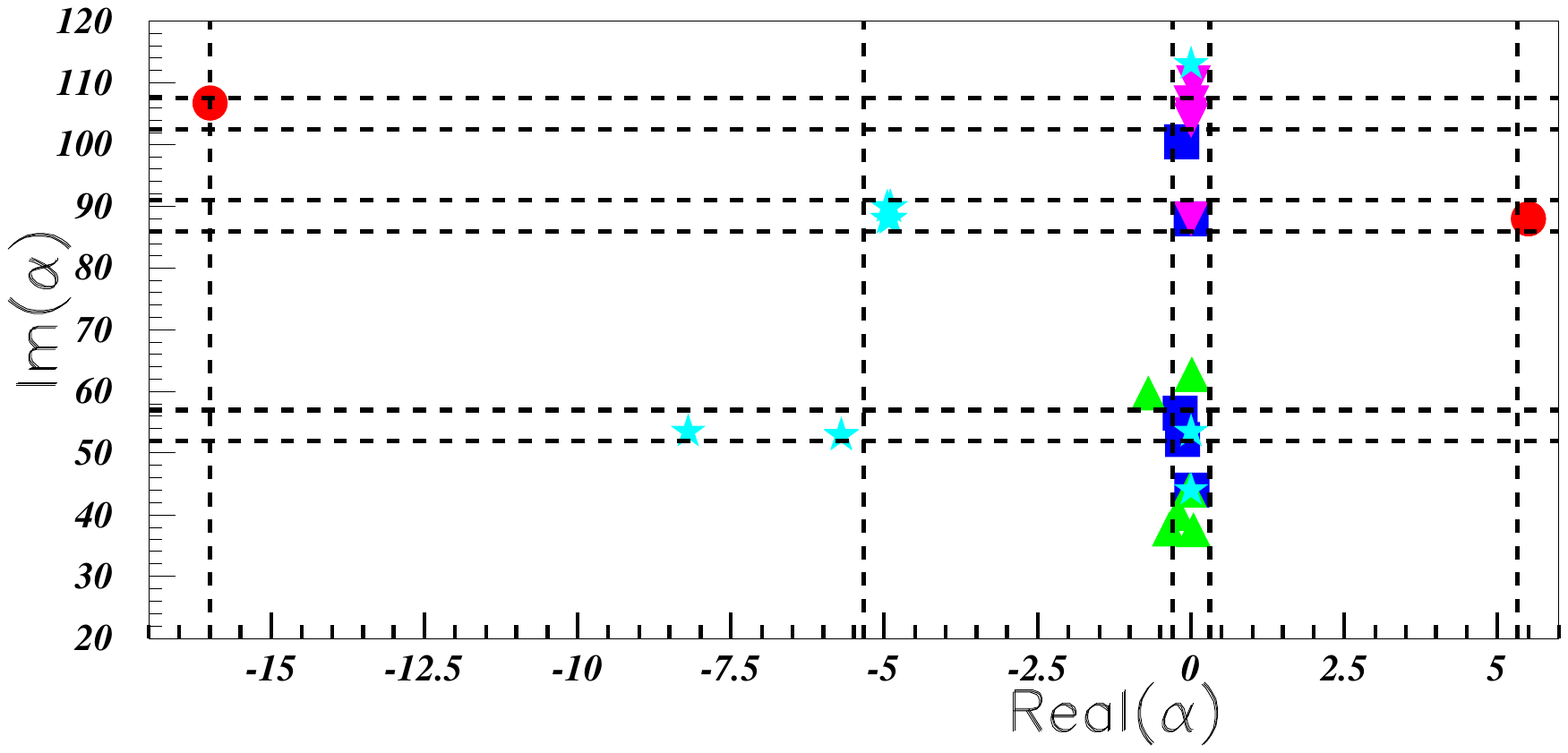}}
\end{figure}

In order to draw the last figure (figure~38), we not only use the $\alpha$ values obtained from the fits performed above through the study of the nuclei masses and nuclei energy levels. We also  use the $\alpha$ values obtained in a previous work \cite{arxiv1104} studying the quark mass ratios, and also the $\alpha$ values obtained in a previous work \cite{frachadron} studying the hadron properties applied to hadron (meson and baryon) spectroscopies. These parameters are given in table IV.

Figure~38 shows the Im($\alpha$) versus Re($\alpha$) plot. Full circles (red on line) show the $\alpha$ values which fit the elementary particle quark and lepton data; there values differ strongly from the others.

Full up side triangles (green on line) show the same for meson data;
the corresponding Re($\alpha$) is stable, close to zero, when the imaginary part varies from 37 up to 63.

Full squares (blue on line) show the same for baryon data; the real part of $\alpha$ is stable close to zero, when the imaginary part moves from 44 up to 101.

Full down side triangles (purple on line) show the same for nuclei mass variation data; here Re($\alpha$)~$\approx$0 is again stable and the imaginary part 
varies a little around 104=2*$\pi$/ln(${\bar \lambda_{1}}$).

Full stars (blue sky on line) show the same for nuclei energy level mass data. 
A great number of marks (seven) are not apparent, since they have the same value: $\alpha$~=~-4.9~+~89*i. The other six marks scatter.
All in all, we have once Re($\alpha)~\approx$-16, once Re($\alpha)~\approx$-8.2, 3 times Re($\alpha)~\approx$, 23 times Re($\alpha)~\approx$0,  and 7 times Re($\alpha)~\approx$5.3.

The marks therefore are not distributed randomly; this is emphasized by dashed lines. There are two distinct Im~($\alpha$) values for excited nuclei energy level masses. These two values correspond to the two distinct values of $\lambda$. At the same values, we observe  Im~($\alpha$) for mesons, quarks, leptons, and data from nuclei mass variation data. All data, except quark, lepton, and data from  nuclei energy level masses have Re~($\alpha$)~$\approx$~0.

In addition to the $\mu$ values, obtained from fits performed on fractal properties on nuclei, the $\mu$ values from elementary particle fits, and hadron fits, reported in table IV, introduce only 
four new values. They are $\mu$~=~0.675 (from quark distribution), 1.08 (unflavored mesons), 1.035 (charmed mesons), and 1.018 ($\Sigma$ and $\Xi$ baryons.)
 
In summary, the analysis of fractal properties of elementary particles masses, hadronic masses, and nuclei masses and energy levels, lead to a rather small number of parameters, much smaller than the very large number of data analysed here (several hundreds).

\section{Conclusion}
We have shown that nuclear physics masses and nuclei energy level masses, as well as elementary particle masses, have often fractal properties and that they verify the discrete-scale invariance properties. Indeed the imaginary part of $\alpha$ is usually more than 18 times larger than the corresponding  real part. We are therefore far from continuous scale invariance. In this model, "the system of the observables obeys scale invariance for specific choices of $\lambda$ (and therefore $\mu$)" \cite{sornette}.

The fractal property of log-log straight line between the masses of several nuclei and their ranks was used to predict some masses of still unobserved nuclei.

The equation used to describe the relative mass increase - equation (3) - allows to reproduce accurately the total spectra of mass variations studied in section II. On the other hand, the fits corresponding to the study of the excited energy levels (section III), reproduce only a part of the used data. Moreover only about 20 data are considered, whereas a much larger number is known. The agreement of theory and data is therefore  weaker concerning this part of the analysis.\\

The analysis of the fractal properties of elementary particles masses, hadronic masses, and nuclei masses and energy levels, leads to a reduced number of parameters, when compared to the large number of masses concerned, masses which therefore appear to be correlated. 

\section{Acknowledgments}
Felix Scholkmann is acknowledged for having pointed out to my attention, the papers studying the particle masses within the Model of Oscillations in a Chain System \cite{mueller}.

\pagestyle{plain}
\begin{table*}[t]
\begin{center}
\caption{Parameters of the nuclear mass ratio and nuclear energy level fits.}
\label{Table III}
\begin{tabular}[t]{c c c c c c c c}
\hline
sect.& fig.& nuclei&$\Omega$&a$_{1}$&$\lambda$&Re($\alpha$)\hspace*{3.mm}Im($\alpha$)&$\mu$\\
\hline
nucl.& 5&A=10&17.5&7.0 10$^{-4}$&1.059&3.8 10$^{-2}$\hspace*{1.mm}110.0&.998\\
mass&7&21&17&2.6 10$^{-4}$&1.061&1.2 10$^{-2}$\hspace*{1.mm}106.8&.999\\
            &9&40&16.7&1.2 10$^{-4}$&1.062&4.0 10$^{-3}$\hspace*{1.mm}104.9&1.\\
  II          & &42&16.7&1.2 10$^{-4}$&1.062&4.0 10$^{-3}$\hspace*{1.mm}104.9&1.\\
              & &38&16.7&1.2 10$^{-4}$&1.062&4.0 10$^{-3}$\hspace*{1.mm}104.9&1.\\
            &12&99&17.0&1.0 10$^{-5}$&1.061&6.4 10$^{-4}$\hspace*{1.mm}106.8&1.\\
          & &100&17.0&2.6 10$^{-5}$&1.061&6.4 10$^{-4}$\hspace*{1.mm}106.8&1.\\
          &14&158&16.5&2.5 10$^{-5}$&1.062&2.7 10$^{-4}$\hspace*{1.mm}104.0&1.\\
           & &192&16.5&1.5 10$^{-5}$&1.062&2.7 10$^{-4}$\hspace*{1.mm}104.0&1.\\      
           & &251&14.0&5.0 10$^{-6}$&1.074&2.7 10$^{-4}$\hspace*{1.mm}88.0&1.\\   
         &16&273&17.0&8.0 10$^{-6}$&1.061&1.6 10$^{-4}$\hspace*{1.mm}106.8&1.\\  
 \hline                    
nucl. &18&$^{11}$C&14.3&0.9&1.072&-4.9 \hspace*{1.mm} 89.8&1.4\\
exci. &  &$^{11}$B&14.0&.62&1.074&-4.9 \hspace*{1.mm} 88.0&1.42\\ 
ener. &21&$^{12}$C&14.3&0.9&1.072&-4.9 \hspace*{1.mm} 89.9&1.4\\ 
  level    &23&$^{13}$C&14.3&0.9&1.072&-4.9 \hspace*{1.mm} 89.9&1.4\\
III   &&$^{13}$N&14.0&.62&1.074&-4.95 \hspace*{1.mm} 88.0&1.42\\
      &&$^{13}$B&14.3&1.0&1.072&-4.95\hspace*{1.mm} 89.8&1.41\\
      &&$^{13}$O&14.0&.62&1.074&-4.9 \hspace*{1.mm} 88.0&1.42\\
    &25&$^{14}$C&7&0.9&1.154&0 \hspace*{1.mm} 44.0&1.\\
    &27&$^{16}$O&8.5&0.9&1.125&-8.2 \hspace*{1.mm} 53.4&2.62\\
    &29&$^{23}$Na&8.45&0.9&1.126&-5.7 \hspace*{1.mm} 53.0&1.96\\
    &32&$^{62}$Ni&8.4&0.9&1.126&-5.7 \hspace*{1.mm} 52.8&1.97\\
  &34&$^{92}$Zr&8.5&3.5 10$^{-6}$&1.125&-7 10$^{-5}$ \hspace*{1.mm}53.4&1.\\
  &36&$^{134}$Ba&18..0&1.5 10$^{-6}$&1.057&-2.6 10$^{-5}$ \hspace*{1.mm}113&1.\\
\hline
\end{tabular}
\end{center}
\end{table*}

\pagestyle{plain}
\samepage
\begin{table*}[t]
\begin{center}
\caption{Parameters of the quark, lepton, and hadronic mass fits. The data corresponding to quark and lepton fits, have been published in \cite{arxiv1104}; the data corresponding to hadronic mass fits, have been published in \cite{frachadron}.}
\label{Table IV}
\begin{tabular}[t]{c c c c c c c c}
\hline
part.& fig.& nuclei&$\Omega$&a$_{1}$&$\lambda$&Re($\alpha$)\hspace*{3.mm}Im($\alpha$)&$\mu$\\
\hline
quark&5&quark&14&1&1.074&5.5\hspace*{3.mm}88&.675\\
lept.&5&leptons&17&1.2&1.061&-16\hspace*{3.mm}106.8&2.56\\
\hline
meson&3&unflavo.&9.5&0.11&1.111&-0.7\hspace*{3.mm}59.7&1.08\\
            & &         &7&.014&1.154&-.02\hspace*{3.mm}44&1.00\\
            &5&strange&10&.018&1.105&.015\hspace*{3.mm}62.8&.998\\
            &7&charmed&6.4&0.03&1.169&-.22\hspace*{3.mm}40.2&1.035\\
            &9&cha.str.&6&.021&1.181&-.35\hspace*{3.mm}37.7&1.06\\
            &13&$c-{\bar c}$&10&.018&1.105&.015\hspace*{3.mm}62.8&.998\\
            &15&$b-{\bar b}$&5.99&.008&1.182&0.04\hspace*{3.mm}37.6&.993\\
\hline
baryon&17&N&14&0.04&1.074&0\hspace*{3.mm}88&1\\
            &20&$\Lambda$&7&0.04&1.154&0.01\hspace*{3.mm}44&.999\\
            &22&$\Sigma$&8.3&.014&1.128&-.14\hspace*{3.mm}52.2&1.017\\
            &24&$\Xi$&9&.022&1.118&-.17\hspace*{3.mm}56.5&1.019\\
            &26&charmed&16&.022&1.064&-.15\hspace*{3.mm}100.5&1.01\\
            &28&$\Xi_{c}$&16&.018&1.064&-.15\hspace*{3.mm}100.5&1.01\\                        
\hline
\end{tabular}
\end{center}
\end{table*}


\begin{thebibliography}{99}
\bibitem{note}The subsections A and B reproduce the model description presented in \cite{frachadron}. They are repeated here for a  complete description.
\end{thebibliography}
\end{document}